\documentclass{iopart}

\usepackage{graphicx}
\usepackage{epsfig}
\usepackage{harvard}
\usepackage{iopams}
\usepackage{lineno}
\usepackage{xspace}
\usepackage{color}

\newcommand{\text}{\rm }
\newcommand{\kev}{ke\kern -0.09em V\xspace}
\newcommand{\gagg}{\ensuremath{g_{\text{a}\gamma\gamma}}\xspace}
\newcommand{\gev}{Ge\kern -0.11em V\xspace} 
\newcommand{\redtext}[1]{{\color{black} #1}}
\newcommand{\degree}{\ensuremath{^\circ}\xspace}
\newcommand{\Rsun}{\ensuremath{R_\odot}\xspace}
\newcommand{\fefifty}{${}^{55}{\rm Fe}$\xspace}
\newcommand{\ev}{e\kern -0.11em V\xspace}

\begin{document}

\title{The X-ray Telescope of CAST}

\author{M~Kuster$^{1,2}$, H~Br\"auninger$^2$, S~Cebri\'an$^{3}$,
  M~Davenport$^4$, C~Elefteriadis$^5$, J~Englhauser$^2$\footnote{on leave},
  H~Fischer$^6$, J~Franz$^6$, P~Friedrich$^2$, R~Hartmann$^{7,8}$,
  F~H~Heinsius$^6$\footnote{on leave}, D~H~H~Hoffmann$^9$,
  G~Hoffmeister$^1$, J N~Joux$^4$, D~Kang$^6$, K~K\"onigsmann$^6$,
  R~Kotthaus$^{10}$, T~Papaevangelou$^4$, C~Lasseur$^4$, A~Lippitsch$^4$,
  G~Lutz$^{8,10}$, J~Morales$^{3}$, A~Rodr\'{i}guez$^{3}$,
  L~Str\"uder$^{8,2}$, J~Vogel$^6$ and K~Zioutas$^{4,11}$}

\ead{\mailto{markus.kuster@cern.ch}}

\address{$^1$ Technische Universit\"at Darmstadt, IKP,
  Schlossgartenstrasse~9, D-64289 Darmstadt, Germany}

\address{$^2$ Max-Planck-Institut f\"ur extraterrestrische Physik,
  Giessenbachstrasse, D-85748 Garching, Germany}

\address{$^{3}$ Laboratorio de F\'{i}sica Nuclear y Altas Energ\'{i}as,
  Universidad de Zaragoza, E-50009 Zaragoza, Spain}

\address{$^4$ European Organization for Nuclear Research (CERN), CH-1211 Gen\`eve 23,
  Switzerland}

\address{$^5$ Aristotle University of Thessaloniki, 54006 Thessaloniki,
  Greece}

\address{$^6$ Universit\"at Freiburg, Physikalisches Institut,
  Herrman-Herder-Strasse 3, D-79104 Freiburg, Germany}

\address{$^7$ PNSensor GmbH, R\"omerstrasse 28, D-80803 M\"unchen, Germany}

\address{$^8$ MPI Halbleiterlabor, Otto-Hahn-Ring 6, D-81739 M\"unchen,
  Germany}

\address{$^9$ Gesellschaft f\"ur Schwerionenforschung, GSI-Darmstadt,
  Plasmaphysik, Planckstr. 1, D-64291 Darmstadt}

\address{$^{10}$ Max-Planck-Institut f\"ur Physik, F\"ohringer Ring 6,
  D-80805 M\"unchen, Germany}

\address{$^{11}$ University of Patras, Patras, Greece}

\begin{abstract}
  The Cern Axion Solar Telescope (CAST) is in operation and taking data
  since 2003. The main objective of the CAST experiment is to search for a
  hypothetical pseudoscalar boson, the axion, which might be produced in
  the core of the sun.
  The basic physics process CAST is based on is the time inverted Primakoff
  effect, by which an axion can be converted into a detectable photon in an
  external electromagnetic field.  The resulting X-ray photons are expected
  to be thermally distributed between $1$ and $7\,\text{\kev}$.  The most
  sensitive detector system of CAST is a pn-CCD detector combined with a
  Wolter I type X-ray mirror system. With the X-ray telescope of CAST a
  background reduction of more than 2 orders of magnitude is achieved, such
  that for the first time the axion photon coupling constant \gagg can be
  probed beyond the best astrophysical constraints
  $\gagg<1\times10^{-10}\,\text{\gev}^{-1}$.
\end{abstract}


\submitto{\NJP}



\section{Introduction}
\begin{figure*}
  \begin{center}
      \centerline{\includegraphics[width=1.\textwidth]{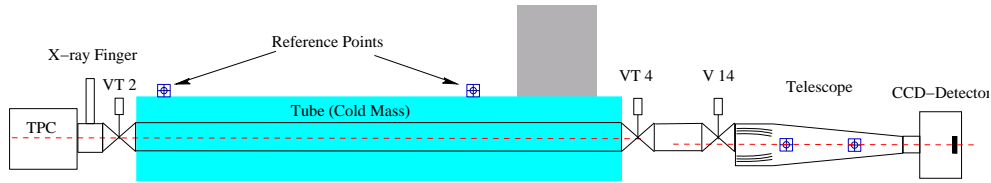}}
  \end{center}
  \caption{\label{fig:experimental-setup} \redtext{Schematic view of the
      experimental setup of CAST. The TPC detector which observes the sun
      during sunset is shown on the left side. On the right side the X-ray
      telescope system is shown. Please note that the Micromegas detector
      and the tracking system are not shown in this picture. The whole
      system is operated at a pressure of $\approx 10^{-6}\,\text{mbar}$}} 
\end{figure*} 
The \redtext{CERN Axion Solar Telescope - CAST,} searches for a
pseudoscalar particle, the axion. \redtext{The axion} was motivated by
theory as a solution of the strong CP problem almost 30 years ago.
\cite{peccei:77a,wilczek:78a,weinberg:78a}. \redtext{One of the most
  important properties of the axion is its coupling to two photons with a
  strength given by the coupling constant \gagg}. This coupling would allow
the production of axions inside the sun via the Primakoff effect
($\gamma\gamma\to\text{a}$) resulting in an axion flux proportional to
$\gagg^2$. The axion energy spectrum would be thermally distributed peaking
at about $3\,\text{keV}$, reflecting the temperature distribution in the
core of the sun \redtext{\cite{sikivie:83a,bibber:89a,andriamonje:07a}}. In
the presence of a transverse magnetic field $B$ of length $L$ solar axions
convert to observable X-rays via the time reversed Primakoff effect with a
probability $P_{a\to\gamma}\propto \gagg^2 (BL)^2$ within the limit of
negligible momentum transfer $q$. In CAST we use a $9.26\,\text{m}$ long
superconducting dipole magnet with an acceptance region of
$14.5\,\text{cm}^2$ providing a $9\,\text{Tesla}$ homogeneous transversal
magnetic field \redtext{to search for solar axions}. \redtext{The magnet is
  mounted on a movable platform which allows to follow the track of the sun
  for about $3\,\text{h}$ per day. On each end of the magnet background
  optimized X-ray detectors are installed, looking for photons from axion
  conversion inside the magnet tube. While the time projection chamber
  (TPC) \cite{autiero:06a} observes the sun during sunset, the X-ray
  telescope and the Micromegas detector
  \cite{andriamonje:06a,charpak:02a,giomataris:96a} are looking for axions
  during sunrise.  \Fref{fig:experimental-setup} shows a side view of the
  CAST magnet and the detector setup. The Micromegas detector which is
  installed next to the X-ray telescope is not shown in this picture.}

For given magnet parameters, the sensitivity of the experiment solely
depends on counting statistics.  The expected count rate from axion to
photon conversion for the X-ray telescope of CAST in the energy range of
$1$ to $7\,\text{keV}$ (spot region) is:
\begin{equation}
  \Phi_{\gamma}\approx1.81\,g_{10}^4\,\text{counts}\,\text{day}^{-1}
\end{equation}
including the effective area of the \redtext{X-ray telescope system}
($g_{10}=\gagg\times10^{10}\,\text{GeV}^{-1}$). \redtext{Taking a mean
  observation time of the sun of $1.5\,\text{h}\,\text{day}^{-1}$ into
  account, the expected signal count rate reduces to $\approx
  0.1\,g_{10}^4\,\text{counts}\,\text{run}^{-1}$.} Thus \redtext{as} in
other rare event experiments background reduction is indispensable to
maximize the sensitivity of the experiment to detect a potential signal.

In this paper we report on the performance, design, and shielding concept
of the X-ray telescope. For a detailed introduction to the CAST experiment
we refer the reader to \citeasnoun{andriamonje:07b}.  First results of CAST
from the 2003 and 2004 data taking period were published by
\citeasnoun{zioutas:05a} and \citeasnoun{andriamonje:07a}. The remainder of
this paper is structured as follows: In
section~\ref{sec:x-ray-mirror-telescope} we give a detailed description of
the design of the X-ray mirror telescope of CAST including the pn-CCD
detector, its calibration, and the alignment of the optical system relative
to the magnet tube. In section~\ref{sec:x-ray-telescope} we present the
long term performance of the system and systematic detector background
studies. We summarize our paper in section~\ref{sec:conclusions-outlook}.

\section{The X-ray Mirror Telescope of CAST}
\label{sec:x-ray-mirror-telescope}

\begin{figure}[t]
  \begin{center}
    \includegraphics[width=0.9\textwidth]{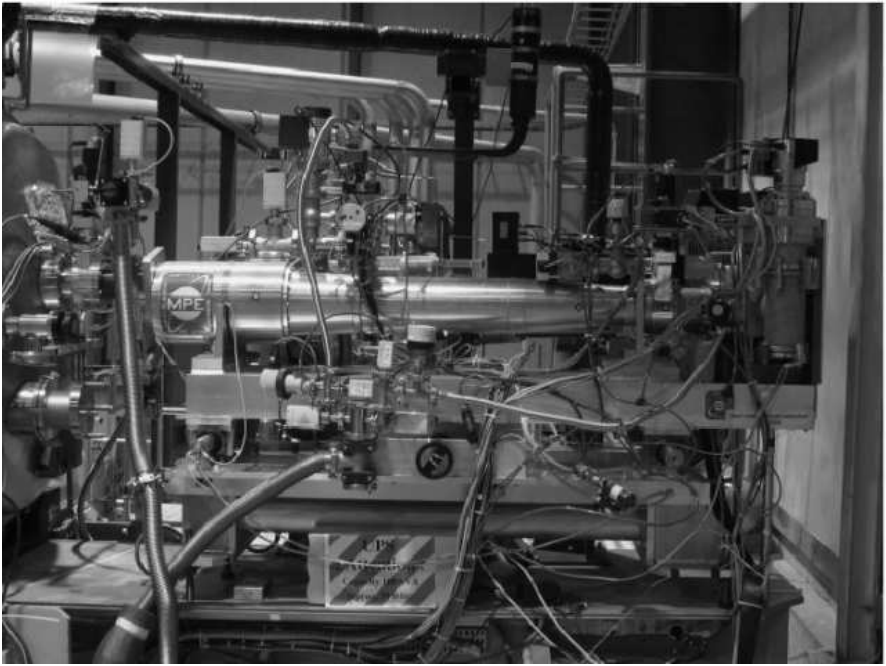}
  \end{center}
  \caption{\label{fig:telescopeandccd} The X-ray telescope consisting of the
    Wolter~I mirror system and a pn-CCD camera are mounted onto the
    superconducting magnet of CAST at CERN. The tube visible in the middle
    of the image houses the mirror module, the magnet bore is to the left,
    and the focal plane pn-CCD detector and its vacuum system is fixed to
    the mirror system to the right.}
\end{figure}

Since the axion to photon conversion inside the magnet tube conserves the
axion energy and momentum in first order approximation, the resulting X-ray
beam would leave the magnet bore with a divergence given by the
\redtext{angular size of the axion producing region of the sun which
  extends from the center of the sun to $\approx 20\%$ of the solar
  radius.}  The resulting X-ray flux can then either be observed directly
with a detector covering the magnet bore, \redtext{as it is the case for
  the Micromegas detector and the TPC of CAST}, or it can be focused with
an X-ray optics onto a focal plane detector with a high spatial resolution.
The advantage of the latter approach is twofold, \redtext{additional
  background suppression by a factor of $\approx 154$ due to the focusing}
of the potential signal from the magnet acceptance area of
$14.5\,\text{cm}^2$ to a small spot of $\approx 9.4 \,\text{mm}^2$.
Furthermore, due to the \redtext{imaging capability an axion image of the
  sun could be acquired and systematic effects can be reduced by measuring
  the background and a potential signal simultaneously, taking} the photon
counts outside the area where the axion signal would be expected into
account. The CAST X-ray telescope is based on the concept of a Wolter~I
mirror optics \cite{wolter:52a} which is a well-known technology used in
X-ray Astronomy (e.g., the Einstein, Exosat, Rosat, Chandra, and XMM-Newton
X-ray observatories) and is a spare module which was originally built for
the X-ray mission ABRIXAS \cite{altmann:98a,egle:98a}. The focal plane
detector is a pn-CCD of the type successfully being used for more than 7
years for ESA's X-ray satellite XMM-Newton \cite{strueder:01a}. The design
and performance of the system will be described in the following sections.

\subsection{The X-ray Mirror System}
\begin{figure*}
  \begin{center}
    \begin{minipage}{0.49\textwidth}
      \centerline{\includegraphics[width=0.957\textwidth]{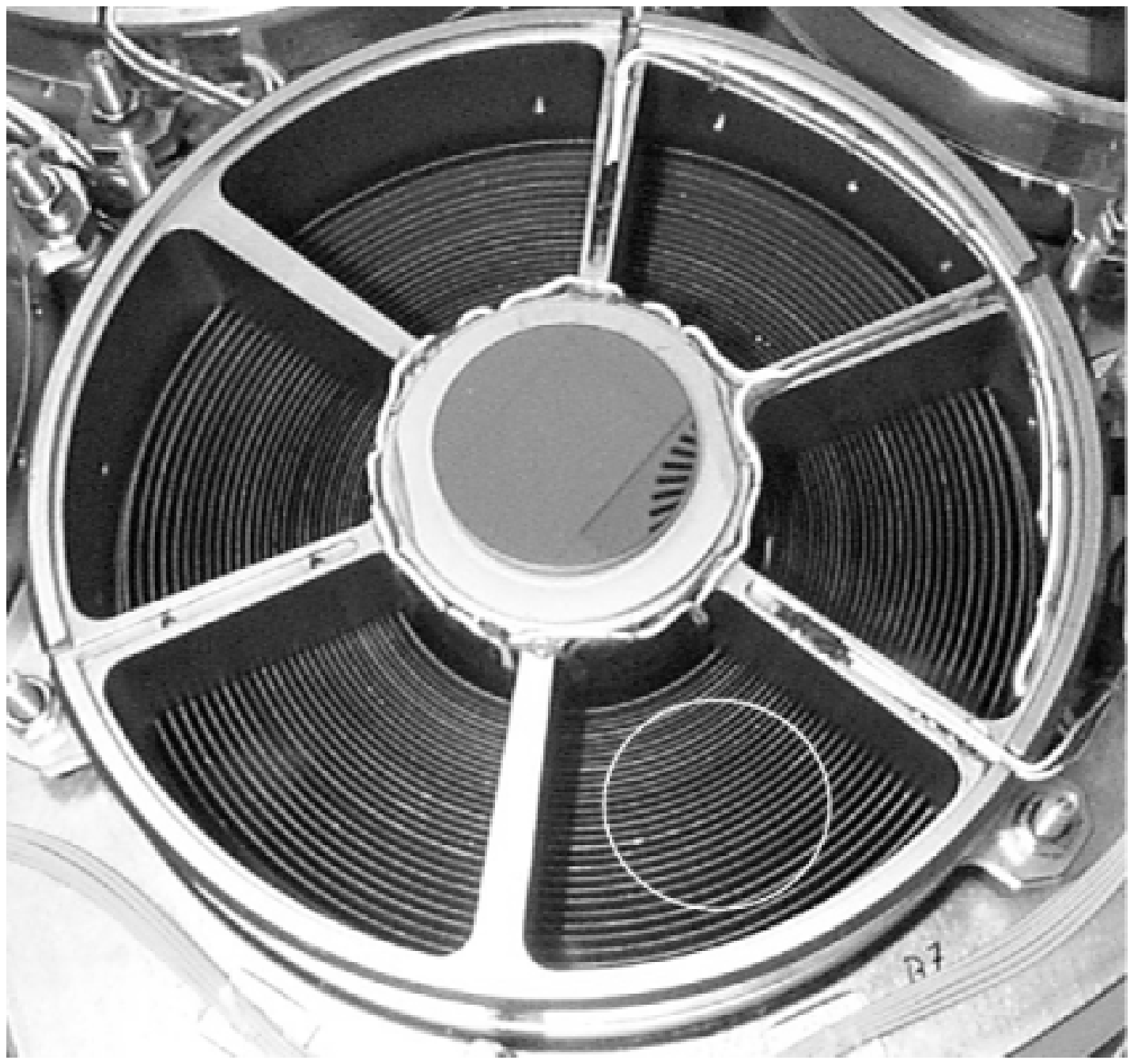}}
    \end{minipage}
    \hfill
    \begin{minipage}{0.49\textwidth}
      \centerline{\includegraphics[width=0.9\textwidth]{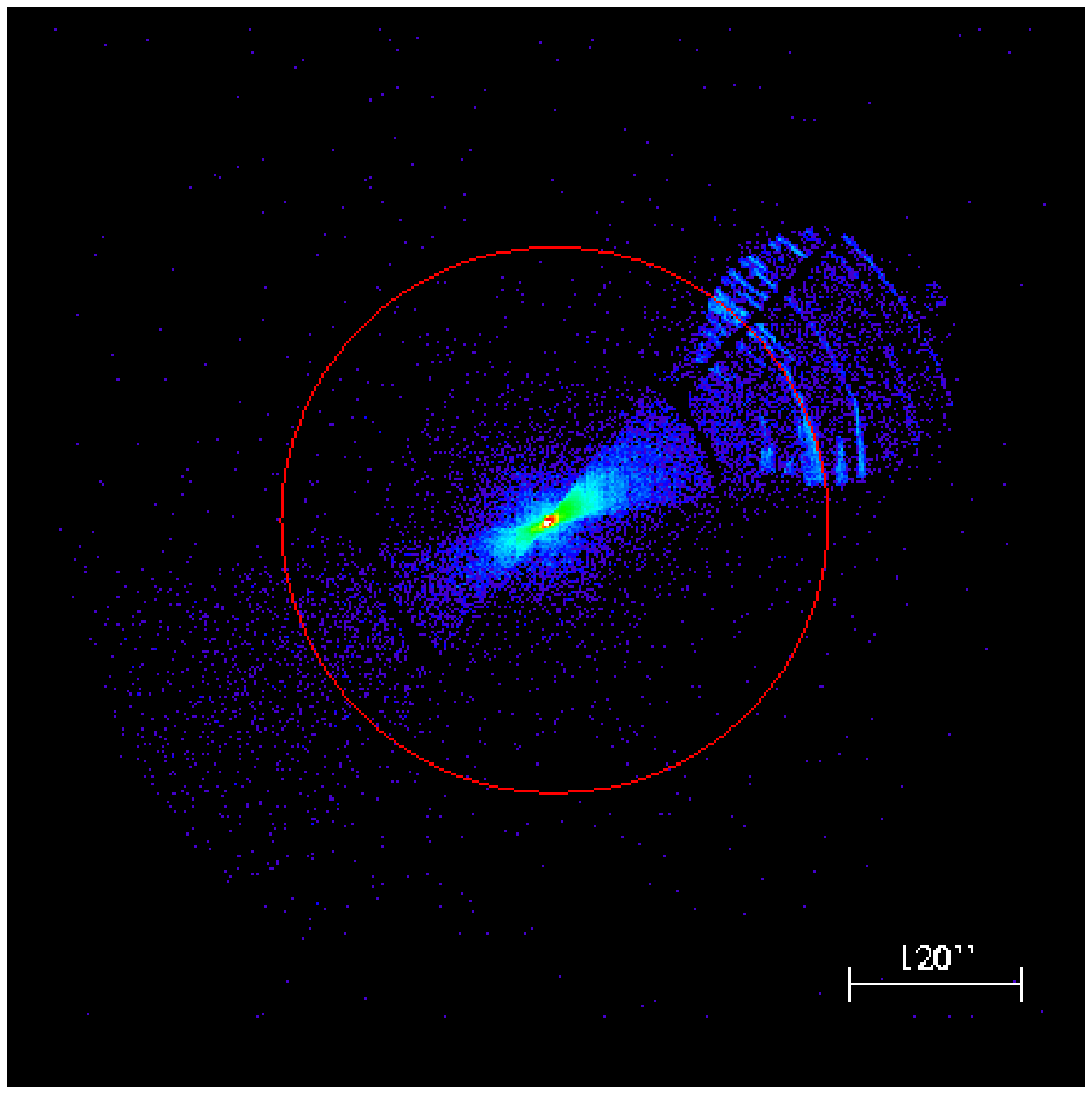}}
    \end{minipage}
  \end{center}
  \caption{\label{fig:mirrormod-psf}Left: Front view of the mirror
    system. The individual mirror shells and the supporting spoke structure
    are shown. One of the six sectors is illuminated through the magnet
    bore, the approximate size of the magnet bore is indicated by the white
    circle. Right: Logarithmic intensity image of a nearly parallel X-ray
    beam measured with one mirror sector at the PANTER test facility at an
    energy of $1.5\,\text{\kev}$. For comparison, the red circle indicates
    the expected spot size of the solar axion signal. Due to the fact that
    the X-ray source is at a finite distance ($d\approx 130\,\text{m}$),
    photons reflected by only one of the parabolic or hyperbolic shaped
    surfaces are apparent in the image (circular shaped region towards the
    top right).}
\end{figure*} 
The CAST X-ray mirror system (see \fref{fig:telescopeandccd}) is a Wolter~I
type telescope consisting of 27 nested, gold coated, and con-focally
arranged parabolic and hyperbolic nickel shells with a focal length of
$1600\,\text{mm}$ \cite{friedrich:98a}. The maximum diameter of the
outermost mirror shell is $163\,\text{mm}$ while the smallest shell has a
diameter of $76\,\text{mm}$. The individual mirror shells are nested in a
spoke structure subdividing the mirror aperture into 6 azimuthal sectors
(see left image of \fref{fig:mirrormod-psf}). Since the diameter of the
bore of the CAST magnet ($d=43\,\text{mm}$) is much smaller than the
diameter of the outermost mirror shell, the telescope is mounted off-axis
such that only one of the six mirror sectors is used for imaging (see
\fref{fig:mirrormod-psf}, the projected size of the magnet bore is
indicated by the white circle). For the application of CAST, the asymmetric
illumination of the mirror aperture has the positive side effect that
shadowing effects caused by the spoke structure do not occur in our setup.

The overall performance of such an X-ray mirror system for a given focal
length mainly depends on two parameters, the effective area and the spatial
resolution, given by the point spread function (PSF). In general, the
effective area for a given mirror coating is a function of the off-axis
angle, the micro-roughness of the mirror surfaces, and the photon energy.
It decreases with increasing micro-roughness, photon incidence angles
(lower reflectivity), and due to geometric effects (vignetting).

The effective area of the CAST mirror system has been predicted by means of
ray-tracing simulations. The algorithm was developed for the ABRIXAS mirror
system and has been adapted to the CAST setup. For a given coating
material, i.e., gold in our case, the ray-tracing simulations return the
effective area and the point-spread function, both as a function of the
photon energy and the off-axis angle. The micro-roughness of the mirror
surfaces has been assumed to be $0.5\,\text{nm}$ \redtext{(rms)} which is a
typical value for the ABRIXAS mirrors. However, the influence of scattering
effects due to the micro-roughness of the reflective surface are almost
negligible.
\begin{figure}
  \begin{minipage}{0.49\textwidth}
    \centerline{\includegraphics[width=1.\columnwidth]{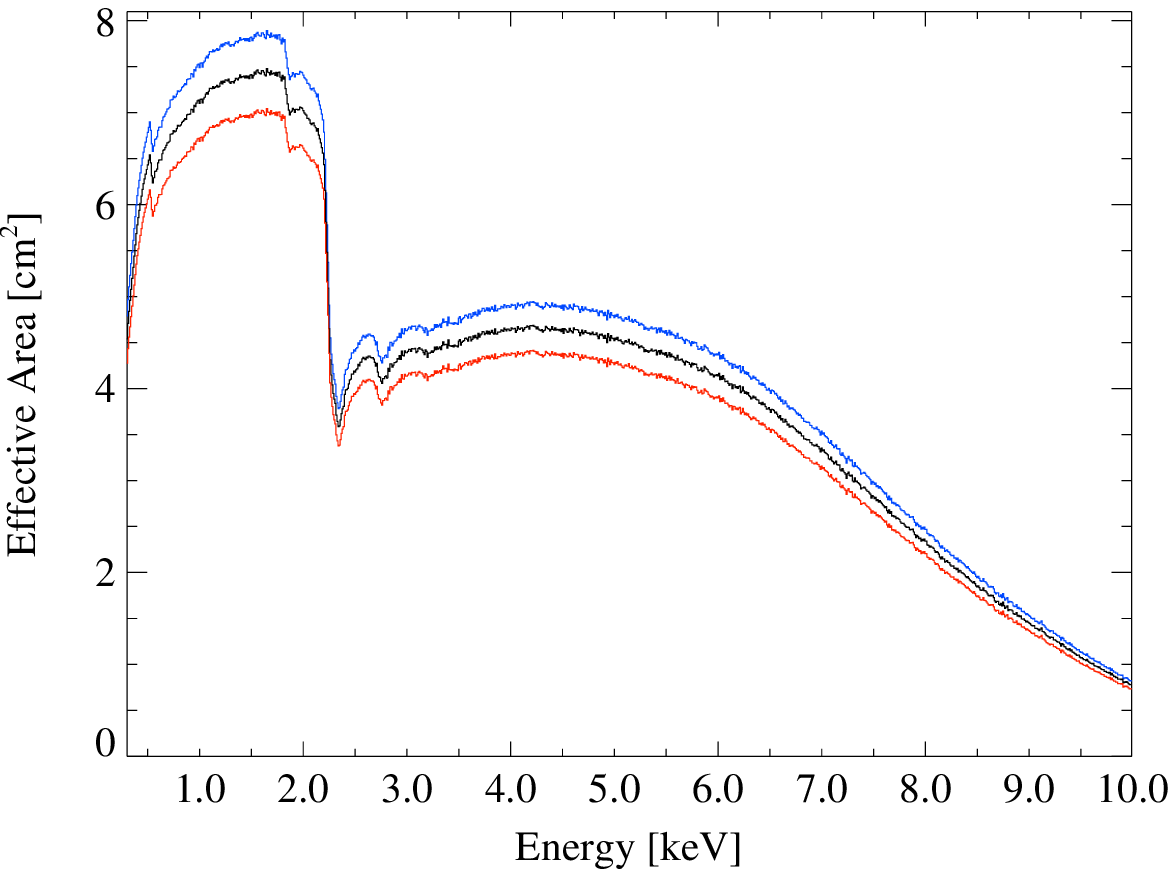}}
  \end{minipage}
  \begin{minipage}{0.49\textwidth}
    \centerline{\includegraphics[width=1.\columnwidth]{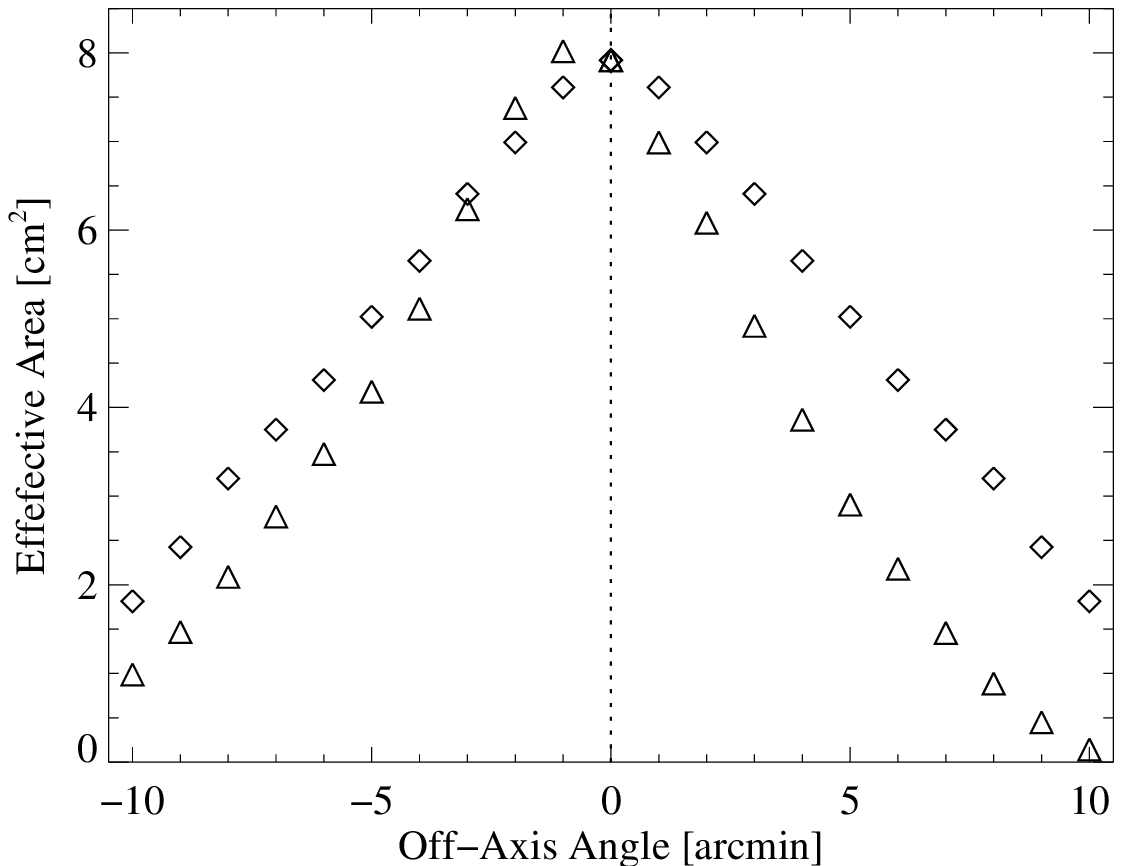}}
  \end{minipage}
  \caption 
  { \label{fig:effarea}Left: Effective area of the X-ray telescope for a
    magnet bore aperture with a diameter of $d=43\,\text{mm}$, given by the
    mirror reflectivity and the quantum efficiency of the pn-CCD.  Three
    different cases are shown: the effective area of the mirror system for
    a point source for the data taking period of 2003 (blue line), the
    effective area for a point source located at an infinite distance for
    the data taking period of 2004 (black line), and for a source of the
    angular size of the axion emission region on the sun for the data
    taking period of 2004 (red line).  Right: Off-axis behavior of the
    effective area of the mirror system only, for the data taking period of
    2003 is shown for two cases: for radial off-axis angles (triangles) and
    tangential off axis angles (diamonds). For a more detailed explanation
    see text.}
\end{figure} 

More important for the actual imaging quality are the figure errors which
influence the point-spread function and the reflectivity of the coating.
Figure errors have not been included in the simulations but have been
determined by X-ray measurements at the PANTER test facility of the
Max-Planck-Institut f\"ur extraterrestrische Physik (MPE)
\cite{freyberg:06a} using monoenergetic X-rays of different energies. The
measurements, performed with full illumination of the mirror system, yield
an on-axis angular resolution of the mirror system of $34.5\,\text{arcsec}$
half energy width (HEW) at $1.5\,\text{keV}$ and $44.9\,\text{arcsec}$ at
$8.0\,\text{keV}$, thus providing an oversampling of a factor of 10 in
spatial resolution compared to the expected size of the ``axion image'' of
the sun. As an example the focal plane image of a point like source
(distance $\approx 130\,\text{m}$) is shown in \fref{fig:mirrormod-psf}.
\redtext{The apparent asymmetry of the focal plane image originates from
  the asymmetric illumination of the mirror system.}

In addition, the energy dependence of the effective area of the mirror
system was measured over a series of distinct energies for each individual
mirror sector. The sector with the best effective area was chosen for CAST.
By combining the simulated effective area with the results of the
calibration measurements, we calculated the on-axis effective area for the
chosen sector and for the energy range important for CAST by interpolation
(see left part of \fref{fig:effarea}). \redtext{In \fref{fig:effarea} the
  effective area for the 2003 data taking period (blue line) is compared to
  the effective area for the 2004 data taking period (black and red line).
  For the 2004 data taking period the effective area is shown for two
  cases:} the effective area for a point source located at infinity
\redtext{(black line)} and for a realistic axion energy and intensity
distribution of the extended solar axion source \redtext{(red line)}. The
overall combined efficiency of the mirror system and the pn-CCD detector
for X-rays from axion conversion varies between $25$ and $46\%$ for the
2004 detector setup depending on the photon energy.  \redtext{In order to
  achieve a better centering of the solar axion spot on the CCD, we had to
  permanently tilt the telescope by $\approx 2\,\text{arcmin}$ relative to
  the axis of the magnet during the data taking period of 2004.  As a
  consequence the efficiency for the data taking period in 2004 is reduced
  compared to 2003.} The right part of \fref{fig:effarea} shows the
simulated radial and tangential (relative to the mirror shell surface)
dependence of the effective area.  According to this picture, the tilt of
the telescope results in a reduction of the effective area by $\approx
10\%$ in absolute value. The radial off-axis behavior shows a slight
asymmetry due to the fact that only one sector of the mirror system is
illuminated, which breaks the radial symmetry of the mirror system.
\begin{figure}
  \begin{minipage}{0.49\textwidth}
    \centerline{\includegraphics[width=1.\columnwidth]{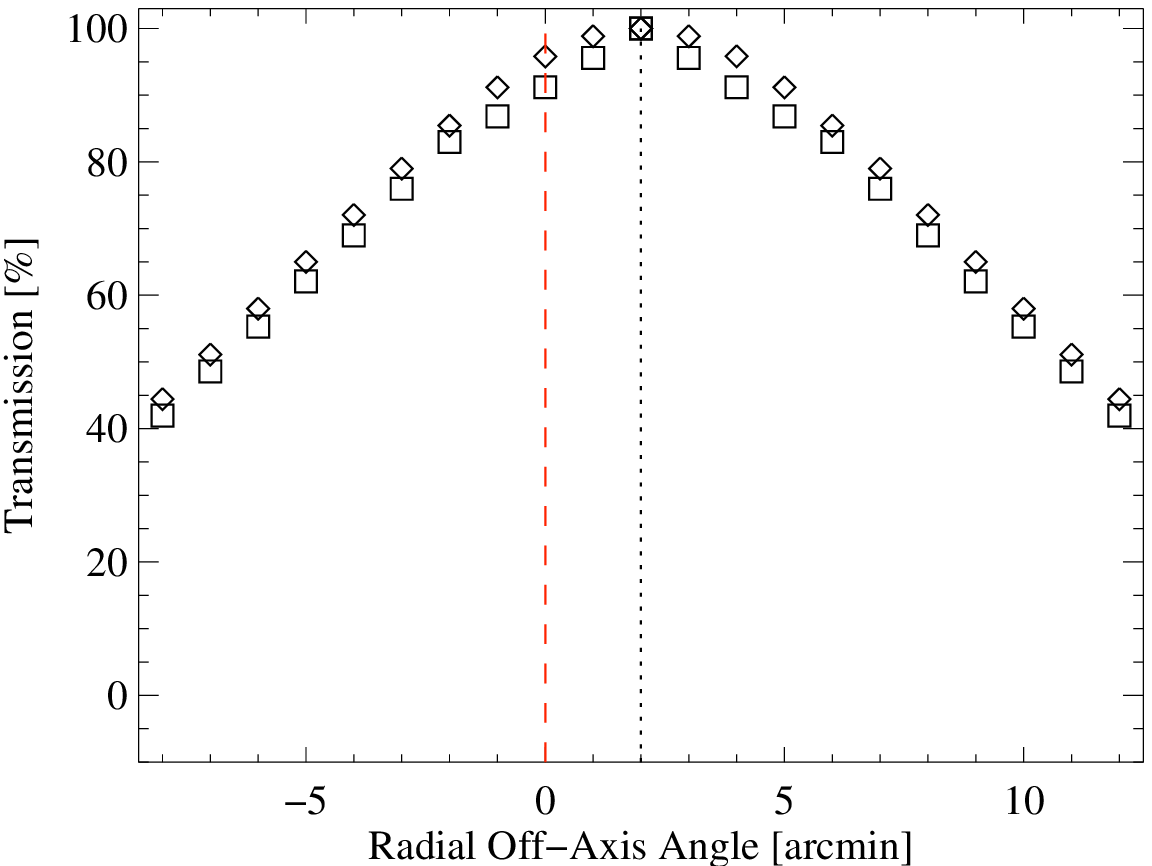}}
  \end{minipage}
  \begin{minipage}{0.49\textwidth}
    \centerline{\includegraphics[width=1.\textwidth]{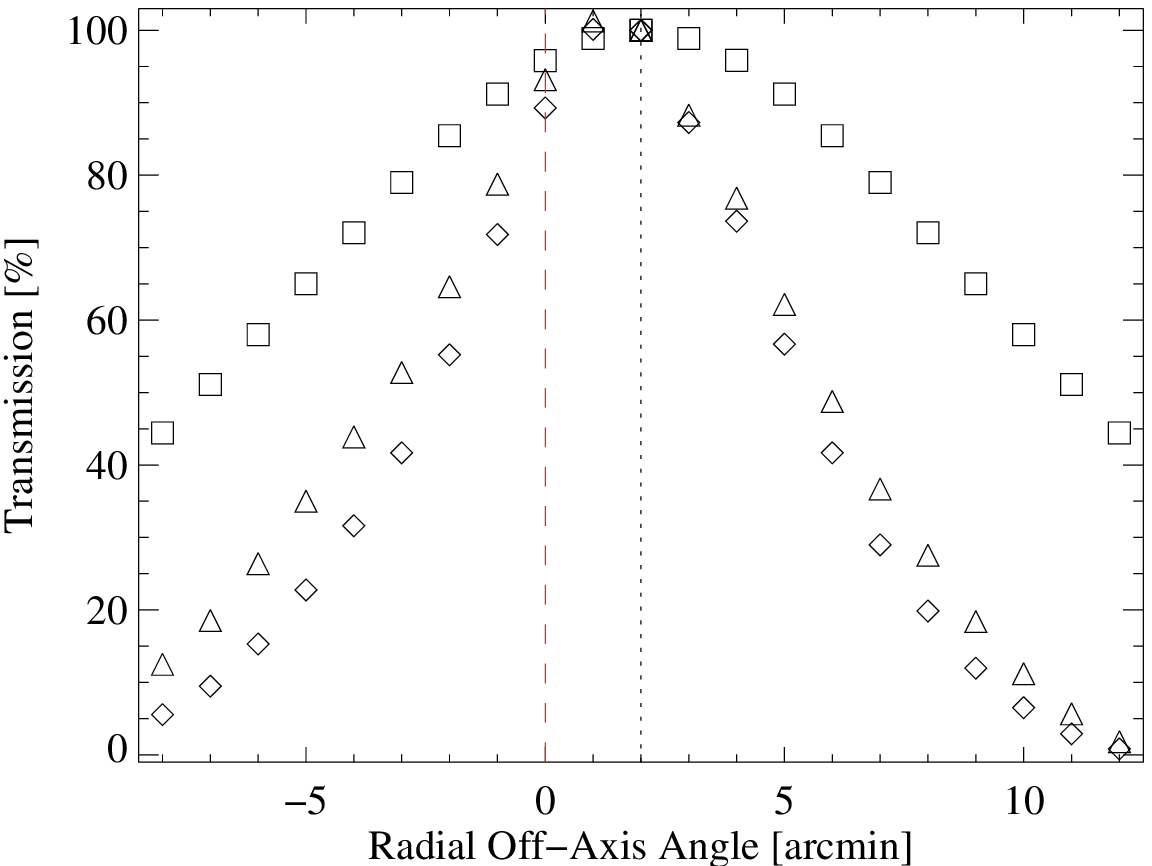}}
  \end{minipage}
  \caption 
  { \label{fig:pointing-vignetting}Left: Vignetting losses due to the
    magnet bore geometry are shown for two cases: a point like source
    (diamonds) and an extended source of the angular size of the axion
    emission region on the sun (rectangles). Right: The transmission of the
    individual components (magnet bore, mirror system) depending on the
    off-axis angle. This includes the vignetting losses due to the magnet
    bore (rectangles), the radial off-axis transmission of the mirror
    module (triangles), and the total transmission of the whole system
    including both components (diamonds, a more detailed explanation is
    given in the text).}
\end{figure} 

In order to estimate geometric effects due to magnet bore and the influence
of the finite size of the axion source, we included the magnet tube
geometry and the shape of the axion emission region on the sun in our
simulations. To simplify matters the beam pipe was assumed to be a
perfectly straight tube with a diameter of $43\,\text{mm}$. The right image
of \fref{fig:pointing-vignetting} shows the combined transmission for an
extended source depending on the off-axis angle for the 2004 data taking
period. The maximum transmission is offset from off-axis angle $0$ due to
the fact that the X-ray telescopes optical axis is slightly tilted
relative to the optical axis of the magnet bore, as mentioned before.

\subsection{The pn-CCD Detector}
\begin{figure*}
  \begin{minipage}{0.59\textwidth}
   \centerline{\includegraphics[width=1.0\textwidth]{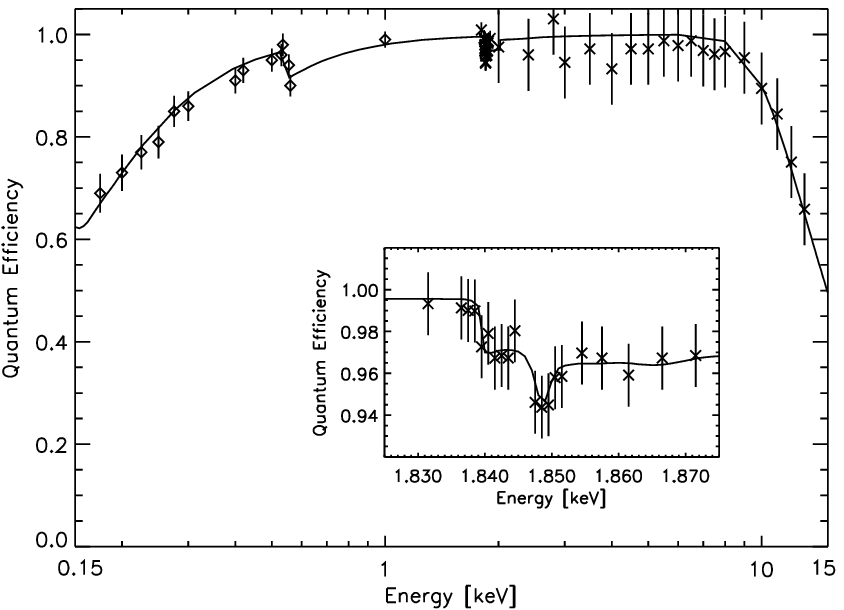}}
  \end{minipage}
  \begin{minipage}{0.40\textwidth}
    \centerline{\includegraphics[width=1.0\textwidth]{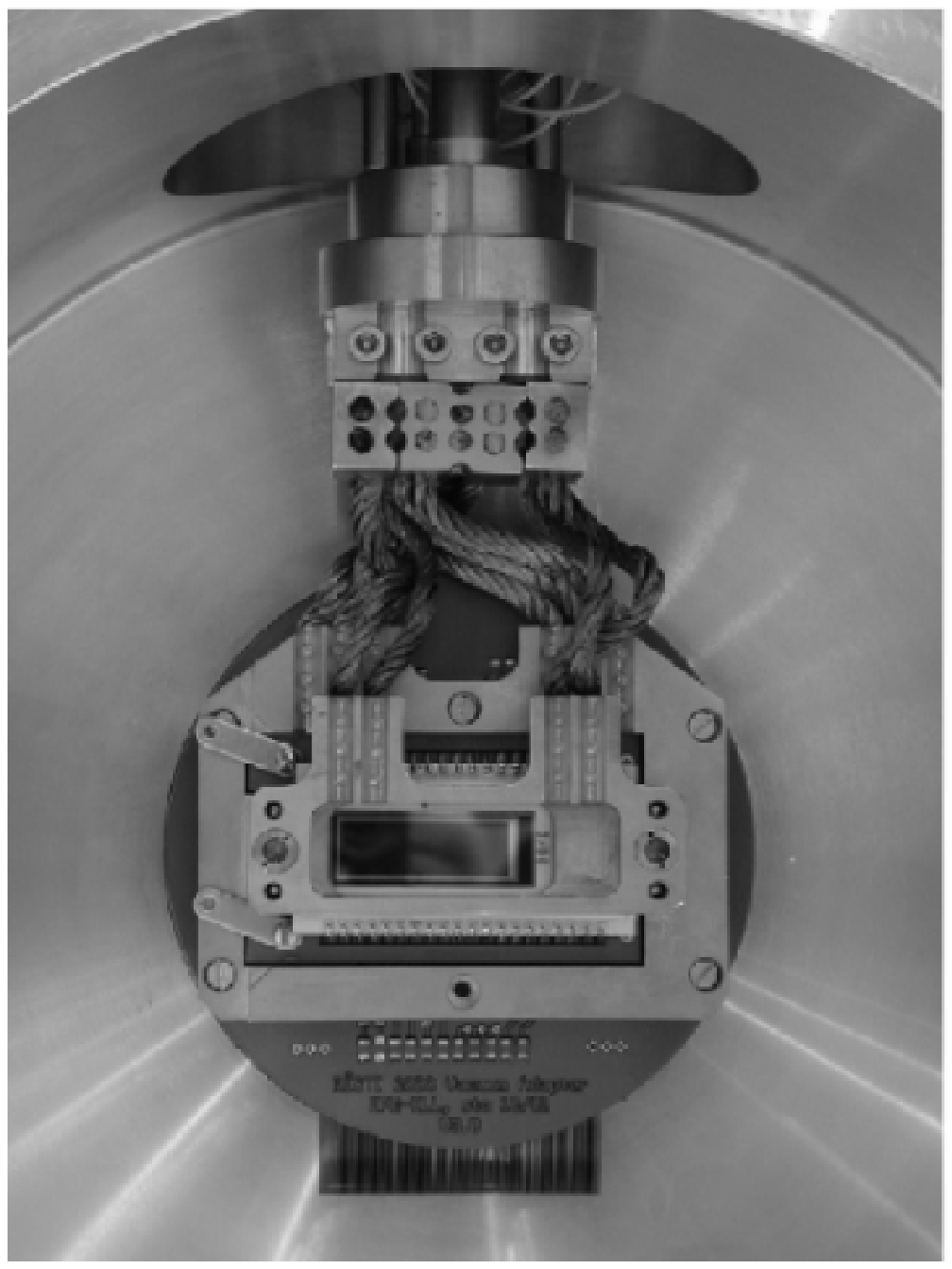}}
  \end{minipage}
    \caption 
    {\label{fig:pnccd} Left: Quantum efficiency (QE) of the fully depleted
      pn-CCD as measured for the EPIC pn-camera of XMM-Newton
      \cite{strueder:01a}. The drop of the QE at $0.53\,\text{\kev}$ is due
      to absorption losses in the $\text{SiO}_2$ passivation layer at the
      detector surface. The inset shows the absorption fine structure close
      to the Si-K edge at $1.84\,\text{\kev}$. The solid line represents a
      detector model fit to the measurements.  Right: The focal plane
      pn-CCD detector inside the CAST telescope. The gold plated cooling
      mask surrounding the rectangular pn-CCD chip (black part in the
      center) is connected to a cold finger of a Stirling cooler on the top
      of the detector chamber via flexible copper leads.  Electrical
      connections to the printed circuit board behind the CCD chip are
      provided via the flex-lead leaving the chamber on the bottom. The
      internal Cu/Pb shield is removed in this picture.}
\end{figure*}

The focal plane detector of the CAST telescope is a $280\,\mu\text{m}$
thick, fully depleted pn-CCD. For a detailed description on the functional
principle and an overview on the general characteristics and concept of
this kind of detector we refer the reader to \cite{strueder:90a} or
\cite{strueder:01a}. The major advantages of such a device are the thick
depletion region and its very thin ($20\,\text{nm}$) and uniform radiation
entrance window on the backside of the chip which results in a quantum
efficiency $\gtrsim 95\%$ in the entire photon energy range of interest
($1$ to $7\,\text{\kev}$) for the solar axion search. The left part of
\fref{fig:pnccd} shows the quantum efficiency measured for a similar
device, the pn-CCD on board of the European X-ray observatory XMM-Newton
\cite{strueder:01a}.

The pn-CCD has a sensitive area of $2.88\,\text{cm}^2$ divided into
$200\times64$ pixels with a size of $150\times150\,\mu\text{m}^2$ each.
This corresponds to an angular resolution of
$19.3\times19.3\,\text{arcsec}^2$ given the focal length of
$1600\,\text{mm}$ for the X-ray optics. The 64 columns of 200 pixels are
read out in $6.1\,\text{msec}$ in parallel followed by an integration
period of $65.7\,\text{msec}$ resulting in a total cycle time of
$71.8\,\text{msec}$.  Since the pn-CCD is operated continously, it is
sensitive to photons all the time and the detector has no dead time.
Although for photons registered during read out, the pixel coordinate in
\redtext{the} shift direction cannot be determined. This results in a
fraction of ``out-of-time'' events of $8.1\%$ assuming a circular intensity
distribution with a diameter of $23\,\text{pixels}\approx3.45\,\text{mm}$
(corresponds to $82.6\%$ encircled axion flux) for the expected solar axion
image. The pn-CCD provides a larger sensitive area than the expected
``axion image'' of the sun.

The operating temperature of the CAST pn-CCD is $-130\degree\,\text{C}$ and
is kept stable with a Stirling cooler system. The thermal coupling between
the cooling system and the pn-CCD is provided by flexible copper leads
connecting the cold finger of the Stirling cooler with the cooling mask of
the pn-CCD chip (see right part of \fref{fig:pnccd}). The detector is
housed inside an aluminum vacuum vessel with a passive shield (removed in
\fref{fig:pnccd}) of typically $10\,\text{mm}$ of oxygen free Cu and more
than $20\,\text{mm}$ of low activity Pb almost free of $^{210}\text{Pb}$ to
reduce the environmental gamma-ray background. An additional lead shield on
the rear side of the detector reduces the gamma background from the
concrete wall of the experimental hall and thus reduces temporal changes of
the background spectrum and level during magnet movement when the distance
between the detector and the wall of the
experimental area changes \cite{kuster:05d}.
\begin{figure*}
  \begin{center}
    \begin{minipage}{0.49\textwidth}
      \centerline{\includegraphics[width=1.\textwidth]{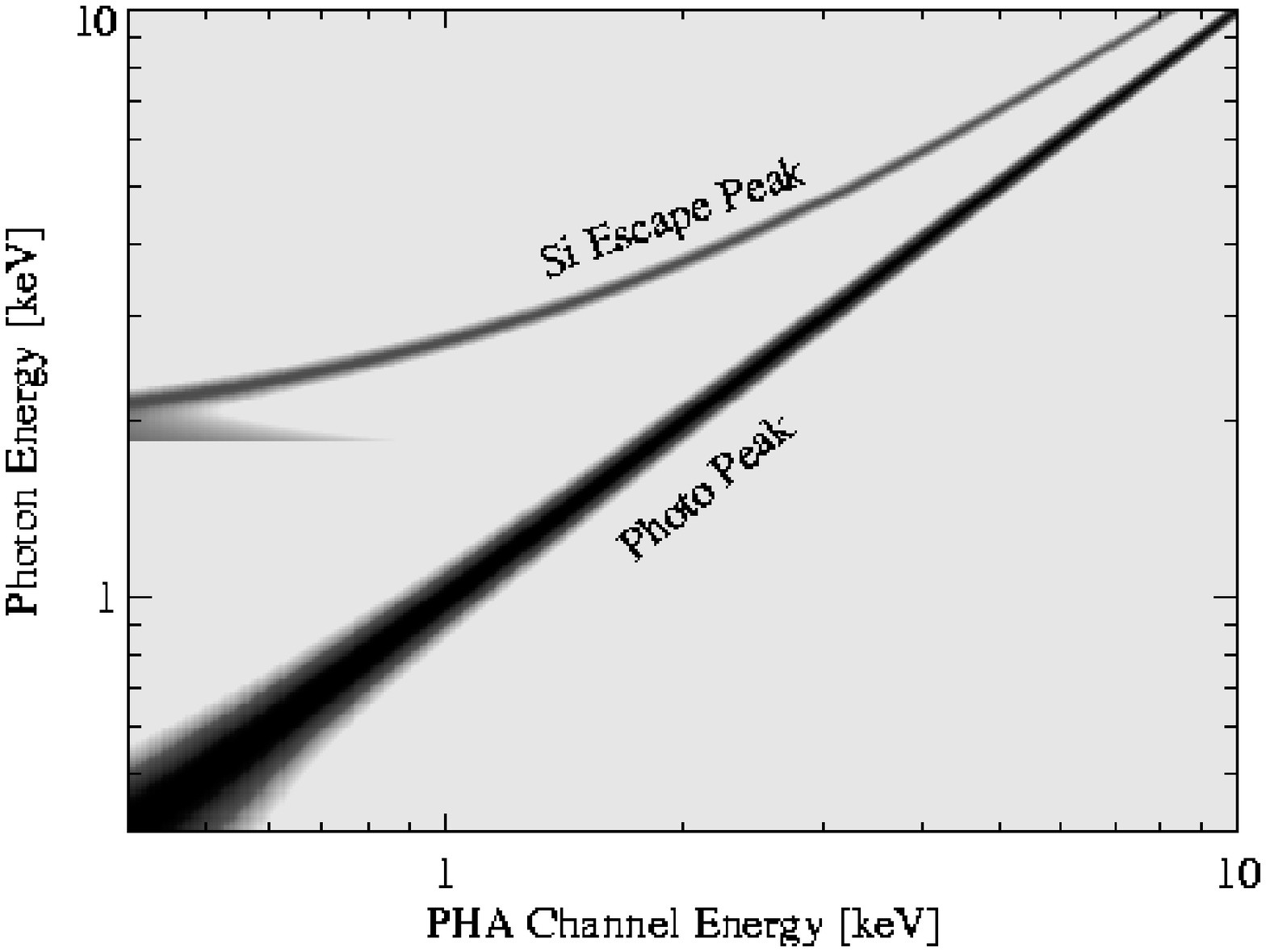}}
    \end{minipage}
    \begin{minipage}{0.49\textwidth}
      \centerline{\includegraphics[width=1.\textwidth]{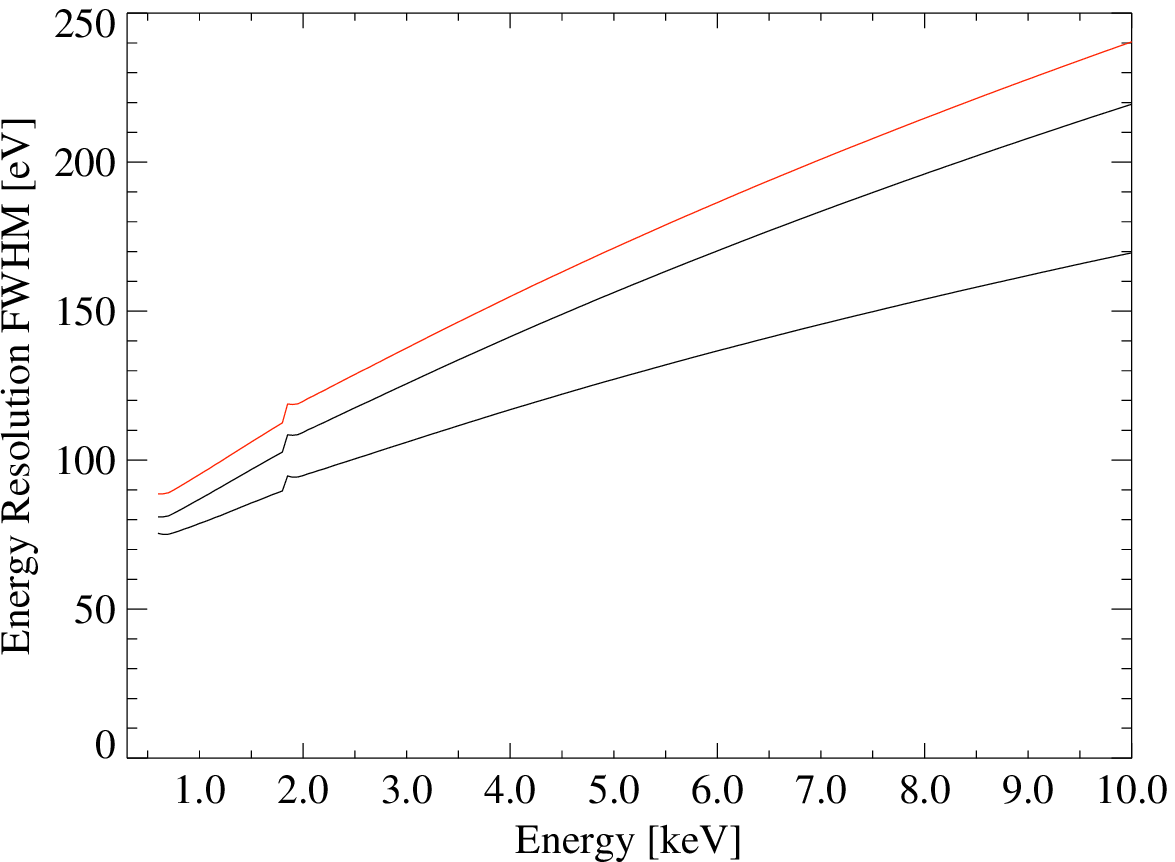}}
    \end{minipage}
  \end{center}
  \caption{\label{fig:response-matrix} Left: Detector response matrix
    of the pn-CCD detector. Apparent are the contributions of the photo
    peak and the $\text{Si-K}\alpha$ escape photons. The width of the
    distribution corresponds to the finite energy resolution of the
    detector. The color scale is logarithmic. Right: Energy resolution of
    the CAST pn-CCD and the EPIC-pn detector of XMM-Newton depending on the
    incident photon energy and different pattern types. From top to bottom:
    The combined energy resolution of the CAST pn-CCD for single and double
    event patterns (see text for a more detailed explanation), the energy
    resolution of the EPIC pn-CCD for double events, and for single events
    only. The XMM-Newton EPIC pn-CCD data is taken from \cite{popp:00a} and
    \cite{haberl:02a}.}
\end{figure*}

Using a multi-target X-ray tube we calibrated the pn-CCD detector in-situ
at CERN. From this data we derived the energy calibration and the detector
redistribution matrix shown in the left part of \fref{fig:response-matrix},
which in combination with the effective area describes the mathematical
relation between the incident binned differential photon spectrum and the
observed binned pulse height spectrum measured by the detector following
the relation:
\begin{equation}
  N_{i} = \sum_{j}R_{ij}\epsilon(E_j)S(E_j)(\Delta E)_j
\end{equation}
Here $N_i$ denotes the number of counts per unit time interval observed in
the energy bin corresponding to the photon energy $E_j$, $\epsilon(E_j)$
the effective area in $\text{cm}^2$, $S(E_j)$ the binned differential
source spectrum in units of
$\text{counts}\,\text{cm}^{-2}\,\text{sec}^{-1}\,\text{keV}^{-1}$, $(\Delta
E)_j$ the finite energy width of the $j$th energy bin, and $R_{ij}$ the
redistribution matrix. The redistribution function includes the
contributions from the photo-peak, the Si escape peak which is expected at
an energy of $1.74\,\text{\kev}$ below the photo peak, and the finite
energy resolution of the detector. Second order effects, like e.g., the
influence of partial events, were not taken into account for the modeling
of the detector response function. We consider these effects of minor
importance for CAST, although they are not negligible for high resolution
X-ray spectroscopy \cite[for a detailed discussion]{kahn:80a,popp:00a}.

Furthermore, we extended the calibration to energies which were not
reachable with the X-ray tube ( $E>9\,\text{\kev}$ ) using X-ray
fluorescent lines apparent from the observed background spectrum (see
Sec.~\ref{sec:long-term-performace} \fref{fig:background-ccd}). Together
with an $^{55}\text{Fe}$ calibration source, these lines also provide a
valuable tool to monitor the long term stability of the energy calibration.
By fitting a sixth order polynomial to these calibration data, we derived
the incident photon energy to detector channel conversion for the energy
range from $0.5\,\text{\kev}$ up to $10\,\text{\kev}$. The energy
resolution of the pn-CCD detector shown in \fref{fig:response-matrix}
depends on the energy of the incident photon and on the pattern type of the
registered event, i.e., whether the charge cloud generated by the incident
photon was registered in one, two, three, or four pixels (single, double,
triple, and quadruple event patterns). Single and double event patterns
contribute in the energy range of $1$--$7\,\text{\kev}$ with a fraction of
$83\%$ and $16.3\%$ to the total number of observed pattern types. The
remaining fraction of $0.7\%$ are triple and quadruple patterns.  To
characterize the energy resolution depending on the incident photon energy
for the CAST pn-CCD we adopted the detector response model which is
actually in use for XMM-Newton EPIC pn-CCD underlying the same physical
detector parameters.  \Fref{fig:response-matrix} shows the resulting energy
resolution of the XMM-Newton EPIC pn-CCD for single and double event
patterns separately, and the resulting energy resolution of the CAST CCD
detector. To model the energy resolution of the CAST detector we combined
both, the noise contributions of single and double event patterns. As
apparent from \fref{fig:response-matrix}, the energy resolution of the CAST
detector is slightly worse compared to the energy resolution of the EPIC
pn-CCD.


\subsection{Telescope Alignment and Pointing Accuracy}
\label{sec:telesc-align}
\begin{figure*}
  \begin{center}
      \centerline{\includegraphics[width=1.\textwidth]{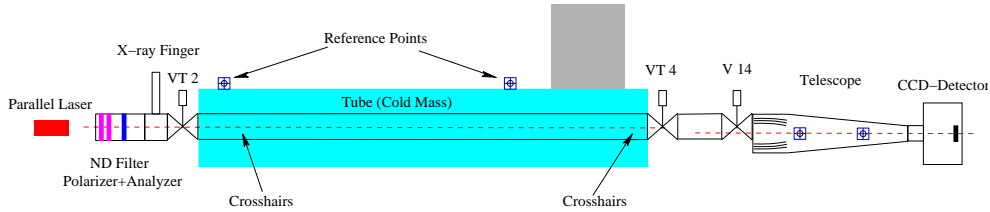}}
  \end{center}
  \caption{\label{fig:alignment-setup} Schematic view of the experimental
    setup of CAST for the alignment of the optical axis of the X-ray
    telescope to the magnet tube axis.}
\end{figure*} 
\begin{figure*}
  \begin{center}
    \begin{minipage}{0.49\textwidth}
      \centerline{\includegraphics[height=0.6\textheight,angle=0]{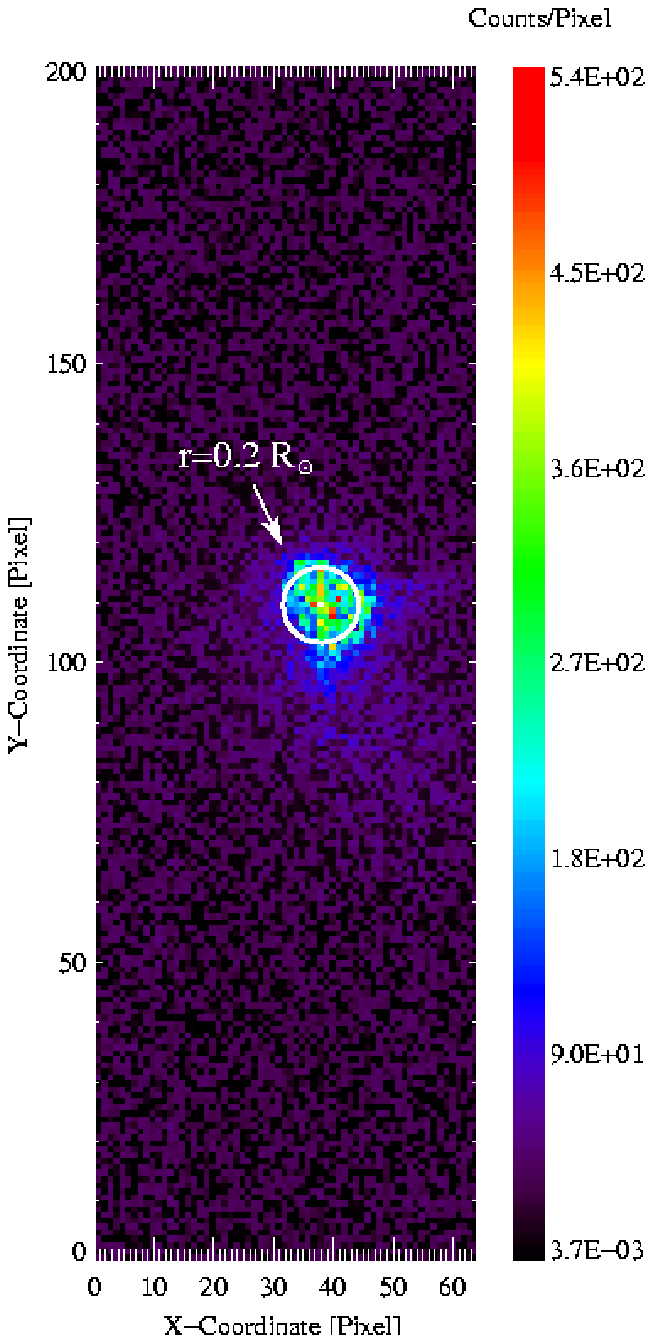}}
    \end{minipage}
    \begin{minipage}{0.49\textwidth}
      \centerline{\includegraphics[height=0.6\textheight,angle=0]{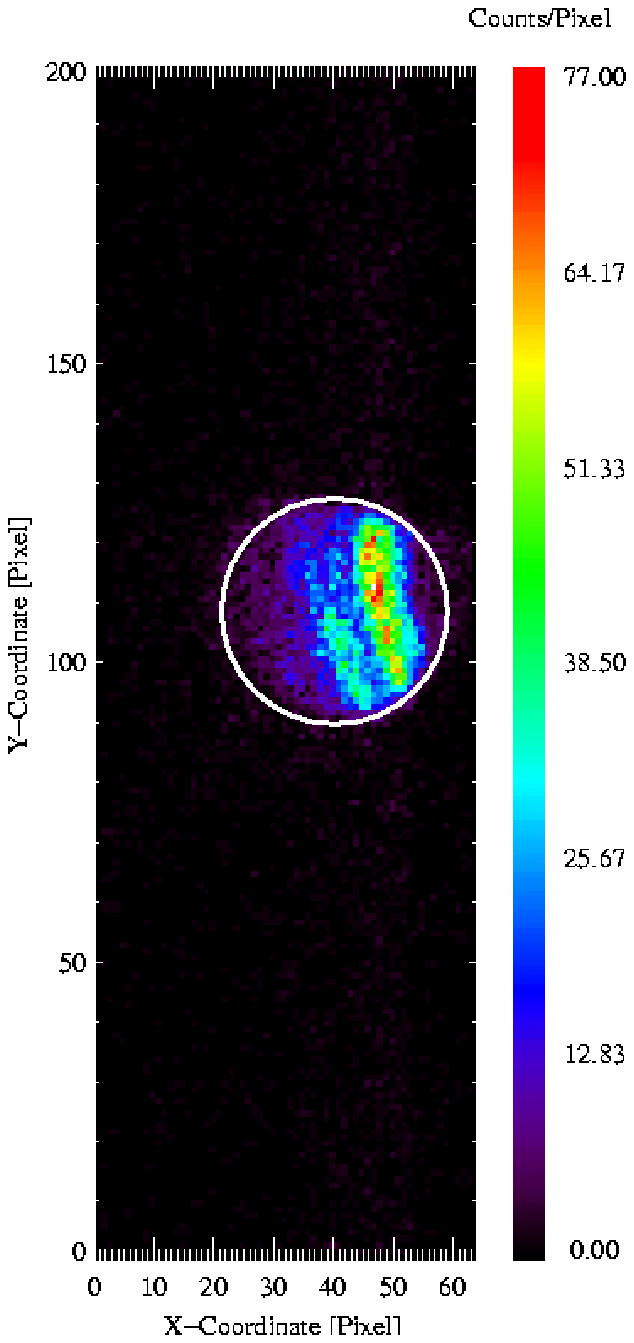}}
    \end{minipage}
  \end{center}
  \caption{\label{fig:alignment-laser-x-ray-spot} Left: The focal plane
    intensity distribution of the parallel laser beam, defining the
    location where an axion signal would be expected. For comparison the
    size of the axion image of the sun is indicated by a circle with a
    radius of $\approx 0.2\Rsun$. Right: The focal plane intensity
    distribution of the $8\,\text{\kev}$ X-ray photons emitted by the
    pyroelectric X-ray source. The circle marks the size of the magnet bore
    projected to the focal plane of the mirror system. The position of the laser spot
    (center of the circle in the image on the left side) has to coincide
    with the center of this circle.}
\end{figure*} 
\begin{figure*}
  \begin{center}
    \begin{minipage}{0.49\textwidth}
      \centerline{\includegraphics[width=1.\textwidth]{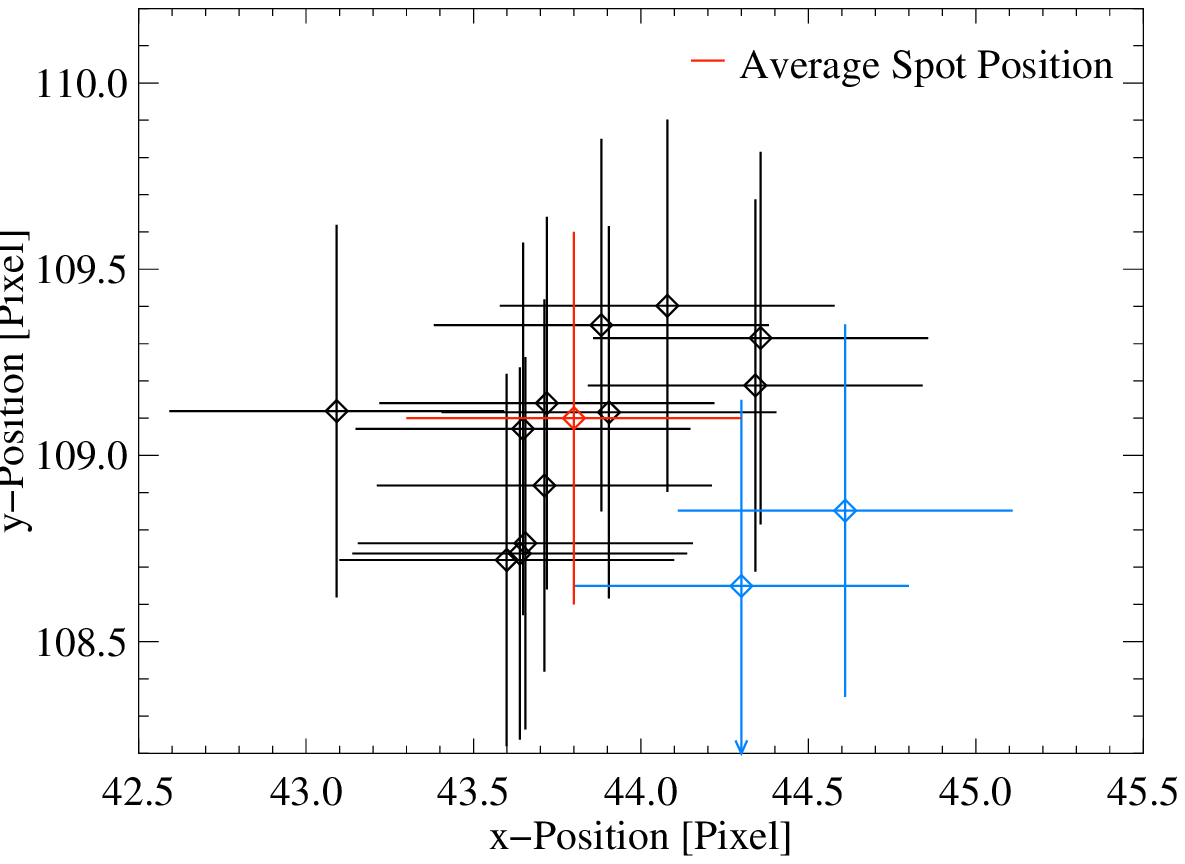}}
    \end{minipage}
    \begin{minipage}{0.49\textwidth}
      \centerline{\includegraphics[width=1.\textwidth]{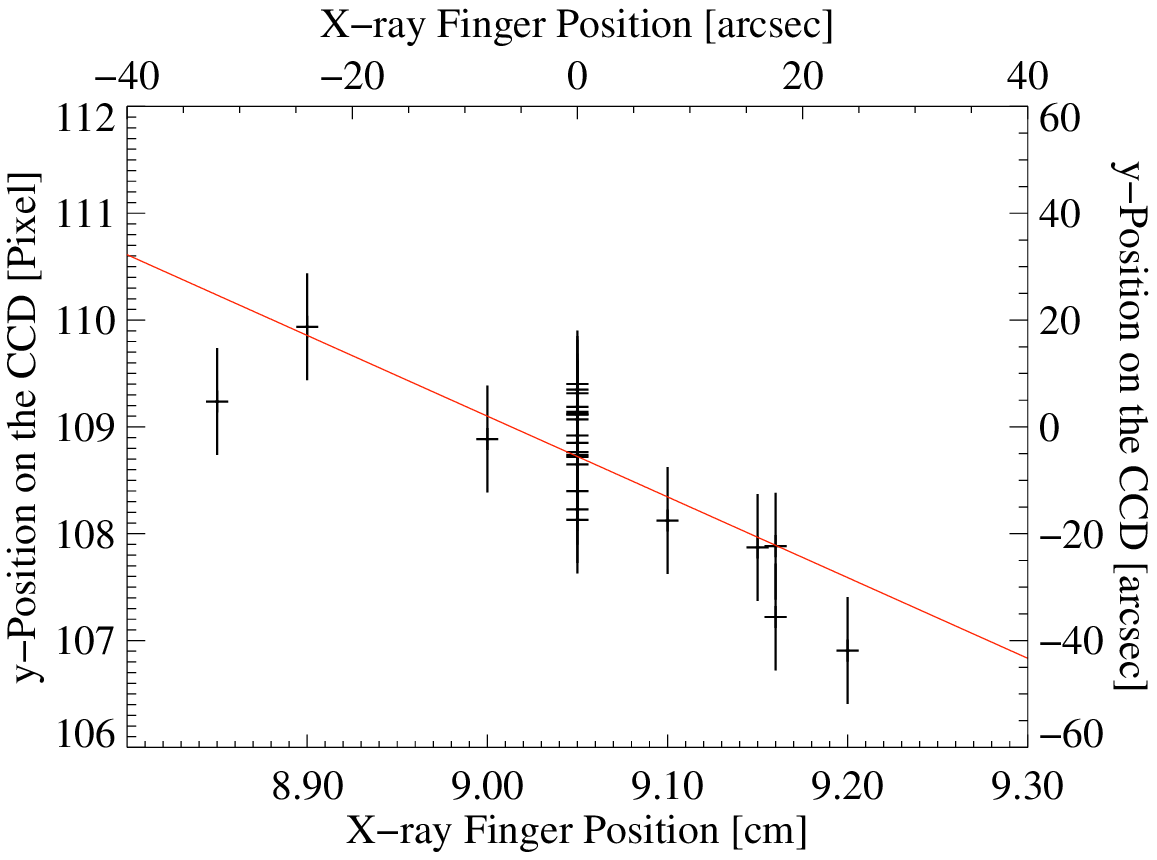}}
    \end{minipage}
  \end{center}
  \caption{\label{fig:alignment-spotposition} Left: Observed location of the
    X-ray spot on the CCD for different magnet orientations. The x- and
    y-position of the X-ray spot on the CCD at the beginning of the data
    taking period in 2004 (black) and at the end of the data taking period
    in 2004 (blue) for different magnet orientations is shown. The red
    cross marks the position averaged over all measurements. Right: Linear
    correlation between transverse X-ray finger position and the X-ray spot
    location as observed with the pn-CCD detector.}
\end{figure*} 
In order to achieve maximum performance of the X-ray mirror system, the
optical axis of the X-ray telescope was aligned to be parallel with the
magnet axis to an angle better than $40\,\text{arcsec}$ using a laser
system providing a parallel beam, shining \redtext{through} the entire
system.  \redtext{For the time of the alignment measurements the laser
  system is installed on the opposite end of the magnet instead of the TPC
  detector. The CCD detector can be replaced by a focusing screen, which
  allows to observe the focal image produced by the parallel laser beam. An
  overview of the experimental setup during the optical alignment is shown
  in \fref{fig:alignment-setup}.}

To be able to monitor the stability of the alignment and the location of
the potential axion image on the CCD detector, for the data taking periods
in 2004, a $\approx 70\,\text{MBq}$ pyroelectric X-ray source emitting
mainly $8\,\text{keV}$ photons was installed on the optical axis of the
system, in front of the TPC detector.  The device labeled ``X-ray Finger''
in \fref{fig:alignment-setup} can be moved in and out of the field of view
of the X-ray telescope.  \redtext{The major advantage of such a device
  compared to a radioactive source is that it can be turned off and
  consequently does neither affect the background level of the TPC nor of
  the X-ray telescope.}  \redtext{Since the source is located at a finite
  distance to the mirror system,} the $8\,\text{\kev}$ photons of the
source are focused $30\,\text{cm}$ behind the CCD.  Therefore, the
observable X-ray image is larger than the focal spot of a parallel X-ray or
laser beam as shown in the right image of
\fref{fig:alignment-laser-x-ray-spot}. The observed intensity distribution
in the focal spot is not uniform since the emission strength depending on
angle of the X-ray finger is non-uniform and thus the focal plane image
represents the emission characteristics of the X-ray finger.  The potential
axion signal is supposed to be located in the center of the circular
envelope of the X-ray spot distribution. After the X-ray finger is aligned,
the position of the X-ray spot can be used to monitor the stability of the
alignment of the X-ray optics and to define the location of the potential
axion image of the sun. The position of the laser spot relative to the
X-ray spot provides an additional consistency check. The size of the focal
spot of the parallel laser filling the magnet aperture should be well
within the expected solar axion spot.

To verify the stability of the alignment, we observed the X-ray spot at the
beginning and towards the end of the data taking period in 2004, during
magnet movement, and for different magnet orientations. The barycenter of
energy of the X-ray spot calculated from each measurement is shown in
\fref{fig:alignment-spotposition} before and after the 2004 data taking
period, and during magnet movement. The measurements yield a stability of
the position of the spot better than $20\,\text{arcsec}\approx
1\,\text{pixel}$ throughout the data taking period of 2004. The overall
pointing accuracy of the CAST helioscope inferred from redundant angular
encoder systems and direct optical observations of the sun is better than
$\approx 1\,\text{arcmin}$, which is perfectly adequate, given the angular
field of view of the magnet bore of $16\,\text{arcmin}$. Since the X-ray
finger was installed in 2004 after the first data taking period in 2003,
the alignment could not be continously monitored during that time. As a
consequence we had to consider a larger and conservatively chosen
extraction region for the potential axion signal on the CCD for the
analysis of the 2003 data \cite{zioutas:05a}. The results of an off-axis
scan, demonstrating the linear correlation of the horizontal position of
the X-ray finger and the location of the observed X-ray spot image on the
CCD is shown in \fref{fig:alignment-spotposition}.


\section{X-ray Telescope Performance}
\label{sec:x-ray-telescope}
\subsection{Long Term Performance}
\label{sec:long-term-performace}
\begin{figure*}
  \begin{center}
      \centerline{\includegraphics[width=1.\textwidth]{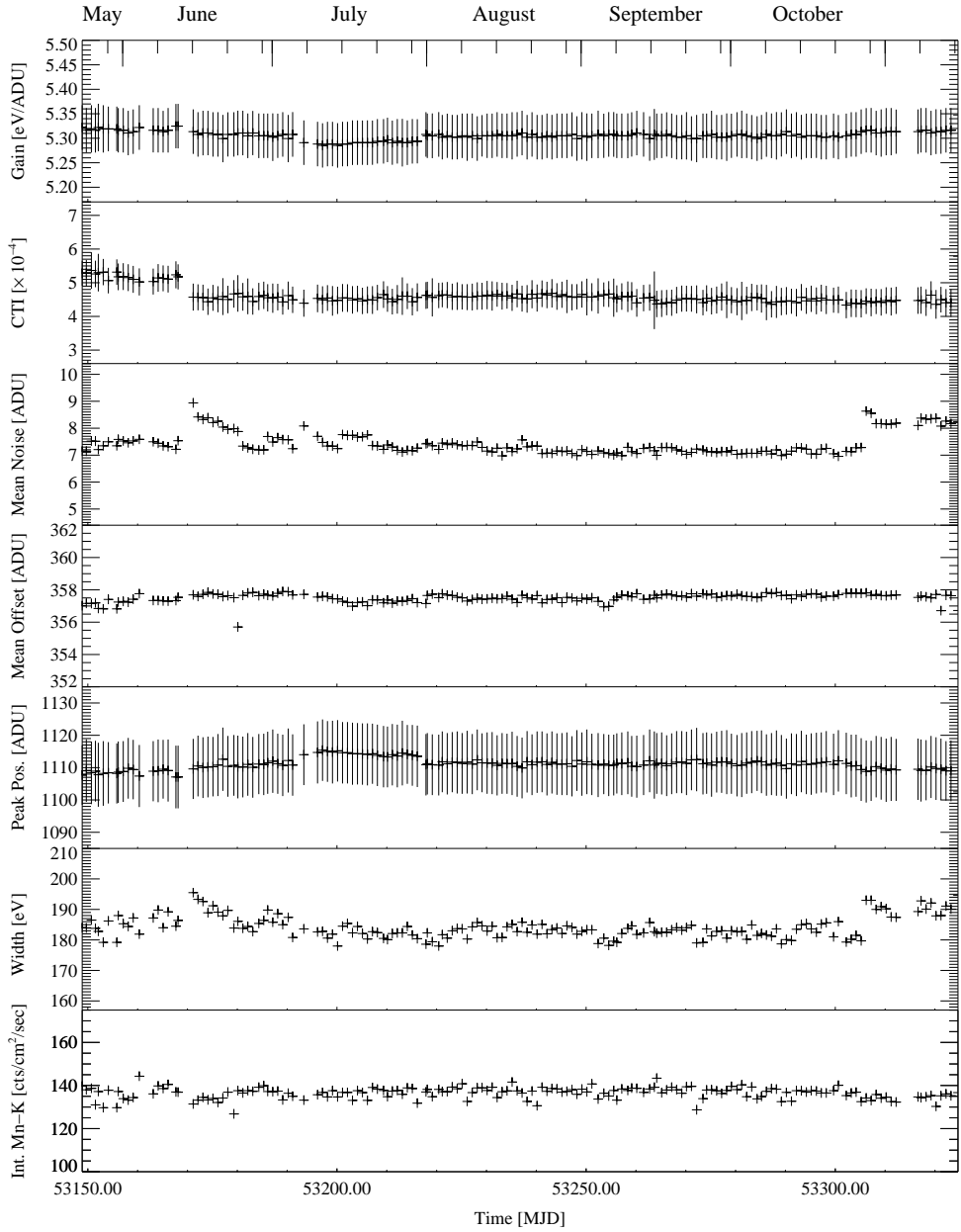}}
  \end{center}
  \caption{\label{fig:hk-summary} Performance of the pn-CCD of the CAST
    telescope during the data taking period of 2004. The results are from
    daily calibration measurements using an $^{55}\text{Fe}$ source. From
    top to bottom are shown: The gain of the detector (i.e., ADU to eV
    conversion), the charge transfer inefficiency (CTI), the mean noise and
    offset averaged over all 12800 pixels, the peak position, the width
    (FWHM), and the intensity of the $\text{Mn-K}\alpha$ line. The time is
    given in UT as ``Modified Julian Day'' (MJD), i.e. $\text{MJD} =
    \text{JD} - 2400000.5$ where JD is the Julian Day.}
\end{figure*} 
A valuable feature of the pn-CCD with integrated front-end readout
electronics is the excellent longterm stability of operating parameters and
performance resulting in homogeneous data sets collected over longer
periods of time. Daily calibration measurements with the CAST pn-CCD using
a flat field illuminating \fefifty source, allow a permanent monitoring of
the performance of the detector. A summary of the most important detector
parameters monitored during the 2004 data taking period is given in
\fref{fig:hk-summary}. Please note, that the errors indicated for the Gain,
CTI, and peak position are dominated by the error of the fitting procedure.
The signal noise averaged over all pixels shows variations which are
correlated to variations observed in the energy resolution of the detector.
Please note that both parameters do not mirror the performance achieved
under controlled laboratory conditions. We assign this to the variable and
sometimes high noise level in the CAST hall which was not designed to be a
low noise experimental area. In no way did the observed degradation affect
the result of the axion search, especially the overall detection
sensitivity for axions of the experiment. All other detector parameters are
stable throughout the data taking period of 2004, similar to the
performance achieved during the 2003 data taking period of CAST.

\subsection{Detector Background}
\label{sec:detector-background}
\begin{figure*}
  \begin{center}
    \begin{minipage}{0.49\textwidth}
      \centerline{\includegraphics[width=1.\textwidth]{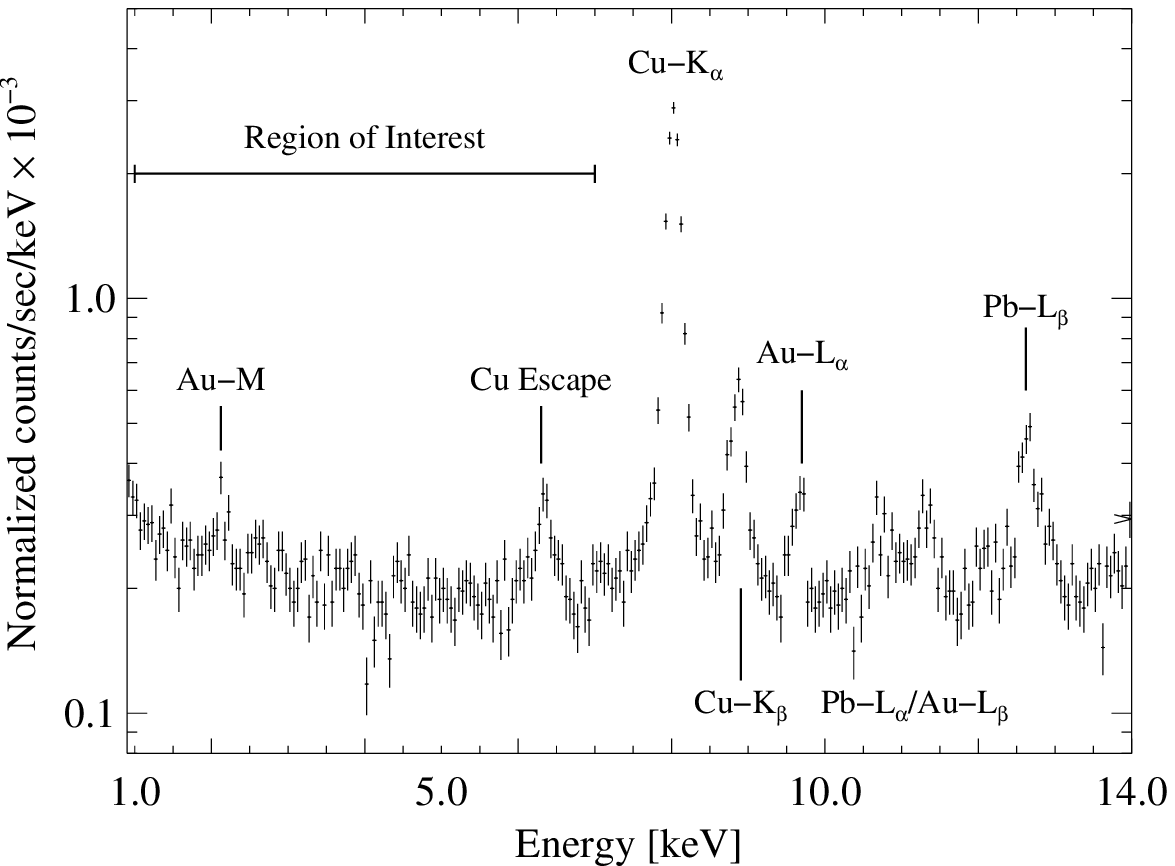}}
    \end{minipage}
    \begin{minipage}{0.49\textwidth}
      \centerline{\includegraphics[width=1.\textwidth]{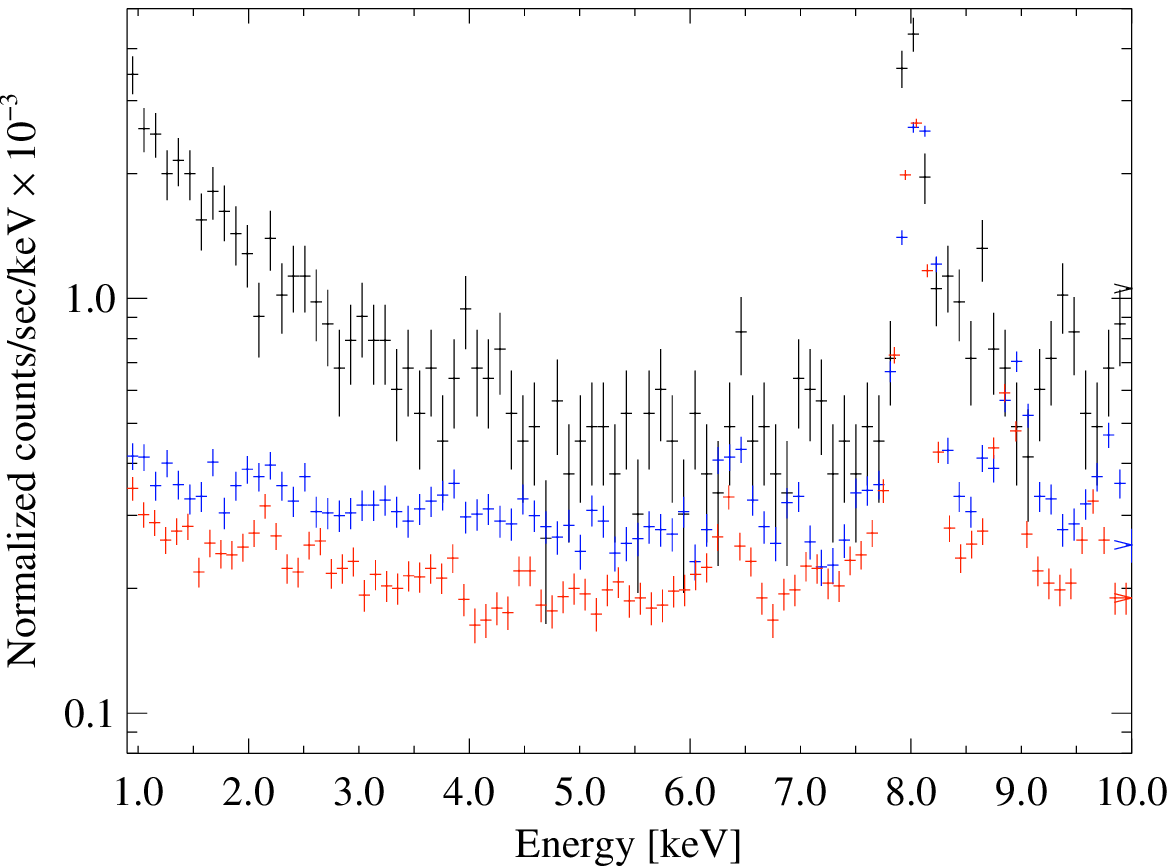}}
    \end{minipage}
  \end{center}
  \caption{\label{fig:background-ccd} Left: Background spectrum observed
    during non-tracking times, i.e., while the CAST magnet was not pointing
    to the sun but under the same operating conditions as during
    observations of the sun \citeaffixed{andriamonje:07a}{see}. The
    fluorescent emission lines apparent in the spectrum are labeled. The
    energy range which is sensitive for axion detection is marked as
    \emph{Region of Interest}. Right: Background spectra observed with the
    pn-CCD under different shielding conditions.  From top to bottom:
    Background observed with the internal copper \redtext{shield (black)
      and with the internal lead and copper shield (blue). The lowest
      spectrum (red) represents the observed background with the final
      shield configuration, which consists (from the outside to the inside)
      of an external lead shield followed by the evacuated detector vessel,
      and the internal internal lead and copper shield.}}
\end{figure*} 
\begin{figure*}
  \begin{center}
    \begin{minipage}{0.49\textwidth}
      \centerline{\includegraphics[height=0.6\textheight,angle=0]{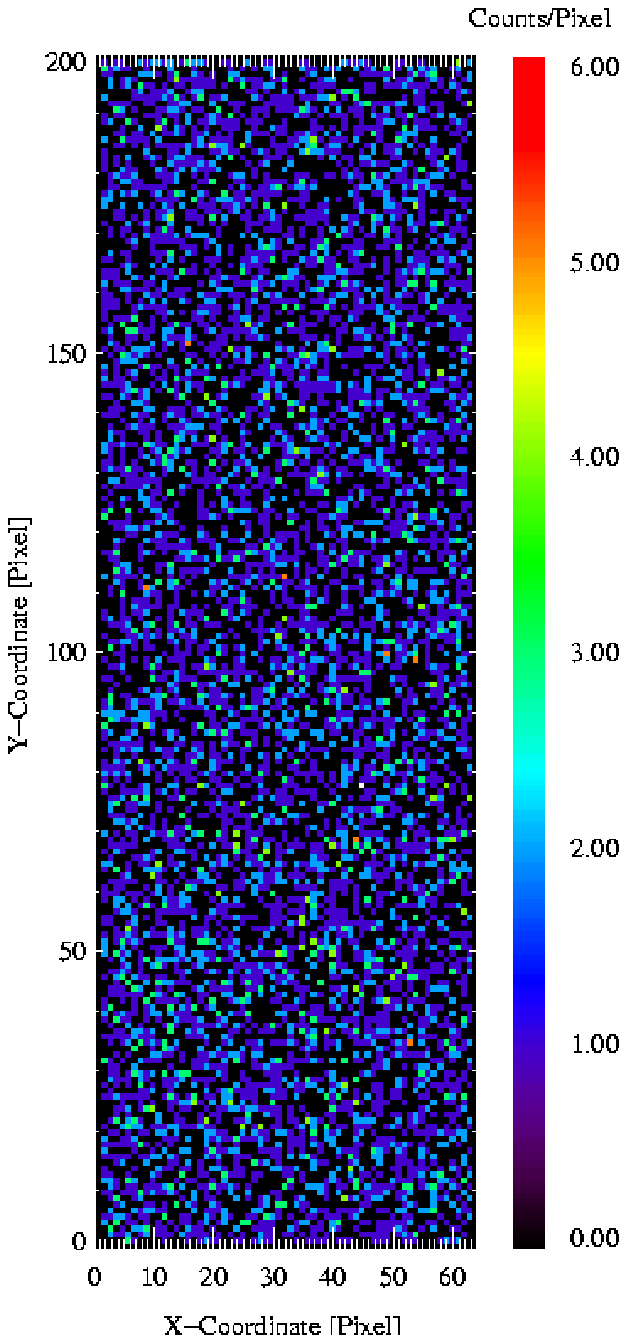}}
    \end{minipage}
    \begin{minipage}{0.49\textwidth}
      \centerline{\includegraphics[height=0.6\textheight,angle=0]{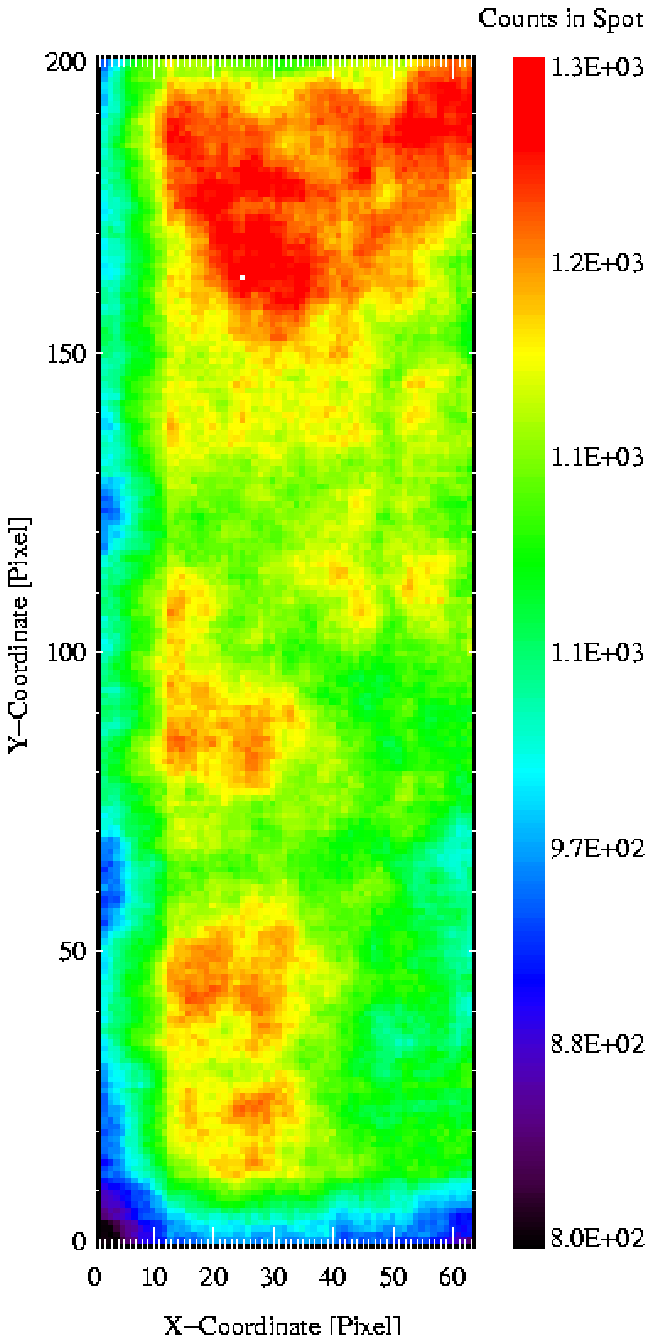}}
    \end{minipage}
  \end{center}
  \caption{\label{fig:ccd-backrgound-distribution} Left: Background spatial
    distribution as observed by the CAST X-ray telescope during the 2004
    data taking period. The intensity is given in counts per pixel and
    integrated over the full observation period of
    $t_{\text{obs}}=6805\,\text{ksec}$. Right: The same data smoothed with
    a circular spot of the size of the expected axion image of the sun.}
\end{figure*} 
In rare event searches which are not free of background, the background
count rate limits the overall sensitivity of the experiment and background
reduction becomes crucial to maximize the sensitivity of the experiment.
\redtext{In general, the detector background can be reduced by a choice of
  radio-pure detector materials, passive or active shielding of the
  detector, by pattern recognition methods, and by minimizing the active
  detector volume by focusing the expected signal to a small area on the
  detector.}  The fact that the CAST experiment is located above surface,
and does not benefit from the shielding effect of the over burden of
underground laboratories, limits the attainable background level. The
strategy to maximize the signal-to-noise level of the X-ray telescope is
therefore twofold: concentrate the potential axion signal on a small area
of the detector and reduce the background by passive shielding as much as
possible. A typical background spectrum measured during the data
acquisition phase in 2004 integrated over the whole CCD sensitive area
which demonstrates the performance of the X-ray telescope is shown in
\fref{fig:background-ccd}.  In the axion sensitive energy range from $1$ to
$7\,\text{keV}$ the resulting mean normalized count rate integrated over
the full detector area is
$(2.21\pm0.02)\times10^{-4}\,\text{counts}\,\text{sec}^{-1}\text{keV}^{-1}$,
corresponding to a mean differential flux of
$(8.00\pm0.07)\times10^{-5}\,\text{counts}\,\text{cm}^{-2}\,\text{sec}^{-1}\,\text{keV}^{-1}$.
The integral background count rate of $0.16\,\text{counts}/\text{hour}$ in
the solar axion spot area ($9.4\,\text{mm}^2$) is remarkably low for an
experiment above surface. The most prominent contributions to the low
energy background apparent from \fref{fig:background-ccd} are the
fluorescent emission lines from material close to the pn-CCD chip, like Cu
(K-photo peaks and escape peak), Au, and Pb. Below $7\,\text{keV}$ the
background is dominated, \redtext{besides the Si escape peak from the Cu
  line}, by an almost flat continuum of predominantly Compton scattered
photons and secondary electrons (for a more detailed explanation see
\citeasnoun{popp:99a} and \citeasnoun{haberl:02a}).
\begin{figure*}
  \begin{center}
    \begin{minipage}{0.49\textwidth}
      \centerline{\includegraphics[width=1.\textwidth]{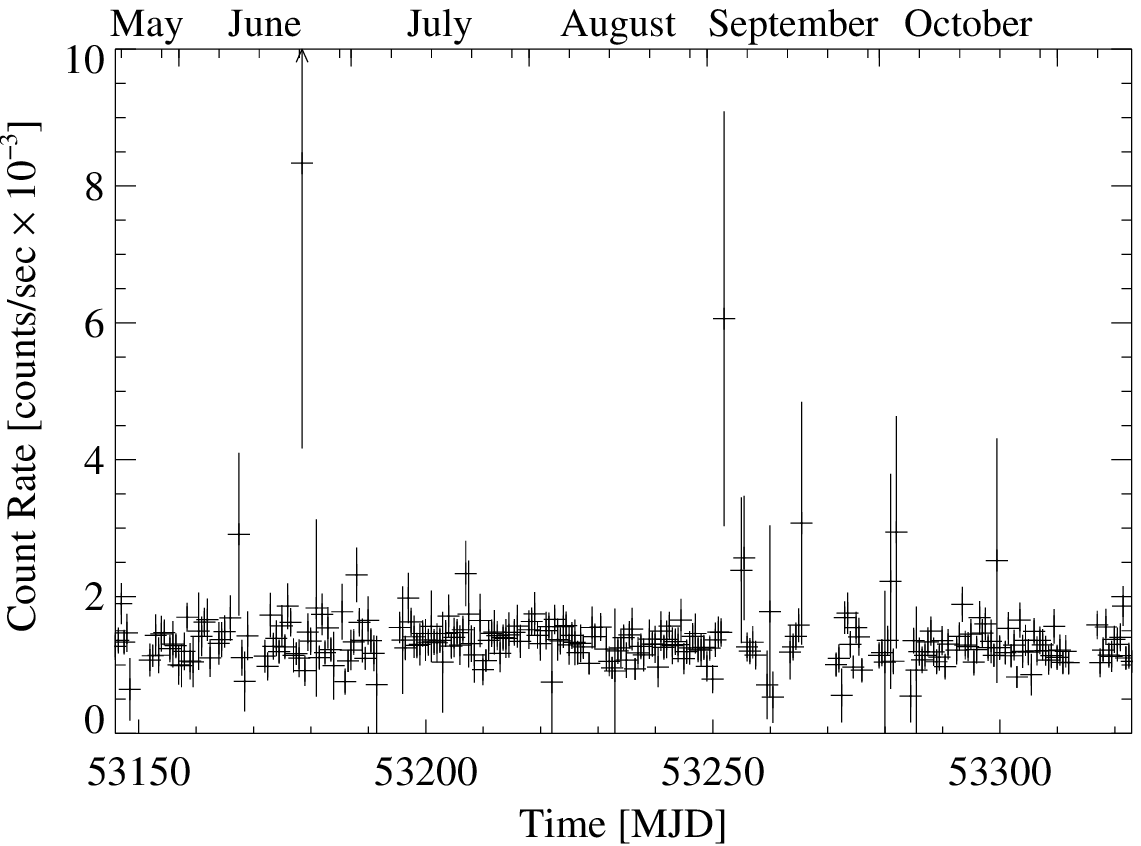}}
    \end{minipage}
    \begin{minipage}{0.49\textwidth}
      \centerline{\includegraphics[width=1.\textwidth]{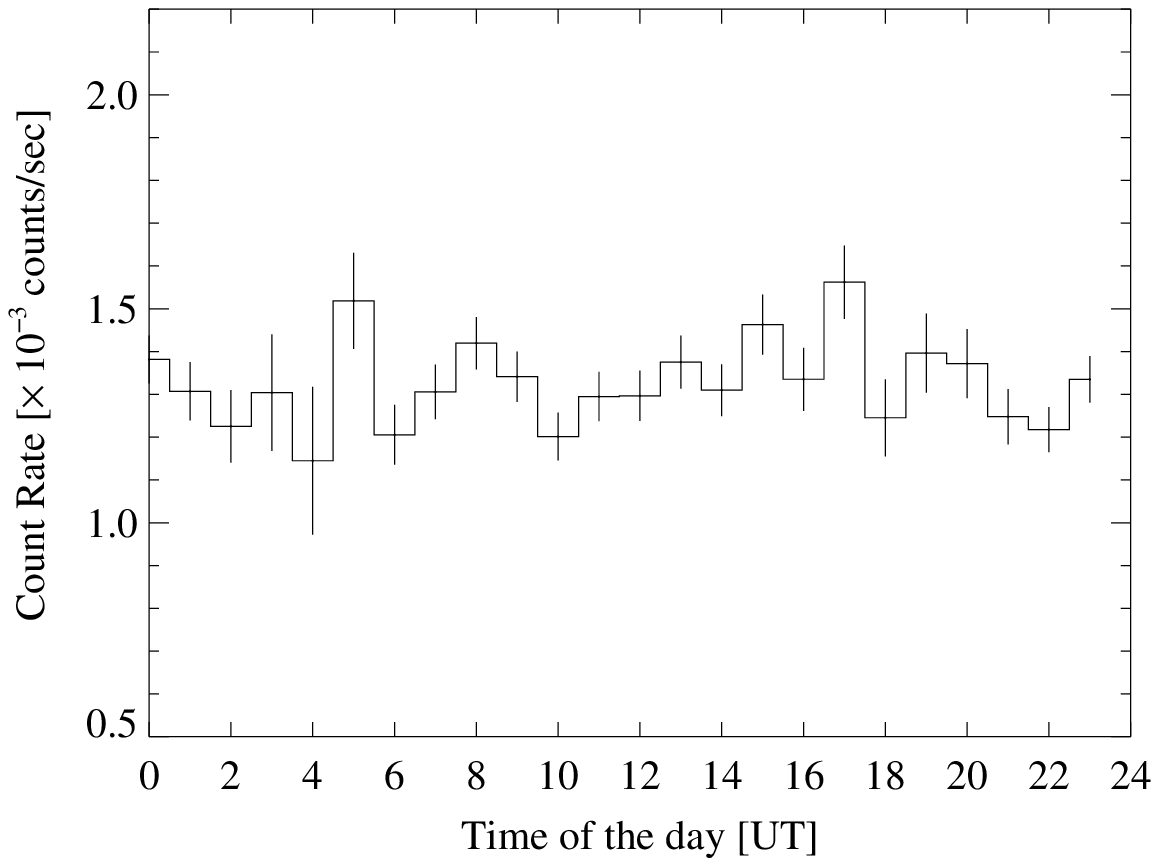}}
    \end{minipage}
  \end{center}
  \caption{\label{fig:background-lightcurve} Left: Time dependence of the
    background in the CAST axion sensitive energy range from $1$ to
    $7\,\text{\kev}$ observed during the data taking period of 2004.  The
    time is given as MJD (UT). Right: Temporal behavior of the background
    during the course of one day in the energy range from $1$ to
    $7\,\text{\kev}$}
\end{figure*} 

The material close to the CCD chip had not been selected for low levels of
radioactive impurities. Nevertheless, samples of all detector components
close to the CCD chip and the pn-CCD chip itself were probed for their
radioactivity in the Canfranc Underground Laboratory of the University of
Zaragoza. Based on these activity measurements, simulations using the
GEANT4 Monte Carlo simulation package were carried out to estimate the
contribution of natural radioactivity to the overall pn-CCD background.
According to the results of the simulations the contribution of natural
radioactivity, e.g. induced by contamination of the detector materials with
$^{238}$U, $^{235}$U, or $^{40}$K can account for at most $\lesssim 33\%$
of the observed background level, whereas about \redtext{$\approx 50\%$} of
the measured background are induced by external gamma-rays originating from
the environment surrounding the detector \cite{rodriguez:07a}.
$^{222}\text{Rn}$ with a half-life of $3.82\,\text{days}$ is usually one of
the strongest sources of natural radioactivity and contributes
significantly to the observed background \cite{heusser:95a}. For the pn-CCD
which is operated in vacuum the contribution of Radon to the total
background is not of importance at the actual level of sensitivity.

\subsection{Background Systematics}
\label{sec:backgr-syst}
A major advantage of the X-ray telescope is the fact that the expected
solar axion image is smaller than the active area of the pn-CCD and thus
background and potential signal can be measured simultaneously during the
observation of the sun by selecting different regions of interest on the
pn-CCD. Nevertheless, we made extensive systematic studies of the observed
background during tracking and non-tracking times and depending on
different operating conditions. \Fref{fig:ccd-backrgound-distribution}
shows the spatial distribution of the events observed with the pn-CCD
during 2004. The data was observed under axion sensitive condition, but
while the CAST magnet was not pointing to the sun.
\Fref{fig:background-lightcurve} shows the corresponding background light
curve (count rate versus time), integrated over the sensitive area of the
pn-CCD. The count rate stays constant at a level of
$(1.32\pm0.04)\times10^{-3}\,\text{counts/sec}$ over the entire data taking
period of 2004 in the energy band of $1$ to $7\,\text{\kev}$. We also
considered the variability of the background on different time scales. The
right image of \fref{fig:background-lightcurve} demonstrates the temporal
behavior of the background count rate during one day, averaged over
different magnet orientations. Selecting different extraction regions on
the CCD, does not affect the results we obtained for the spectral
distribution and the temporal behavior of the background.

A statistical analysis of the background count rate measured in 2004 while
the telescope was pointing at the sun reveals a gradual decrease with time
to a level of about $80\%$ of the rate observed at the beginning of the
2004 data taking. Such a temporal behavior could be due to the systematical
change of the detector location while following the solar azimuth at
sunrise for a period of $176$ days. The detector moves progressively away
from the concrete wall of the experimental hall which according to MC
simulation results could be a significant background source. During the
4-months data taking period in 2003 when the detector shielding was less
hermetic and the background rate was higher by a factor of $1.5$ no such
change of background rate was observed.

\begin{figure}[t]
  \begin{center}
    \begin{minipage}{0.49\textwidth}
      \centerline{\includegraphics[width=1.\textwidth]{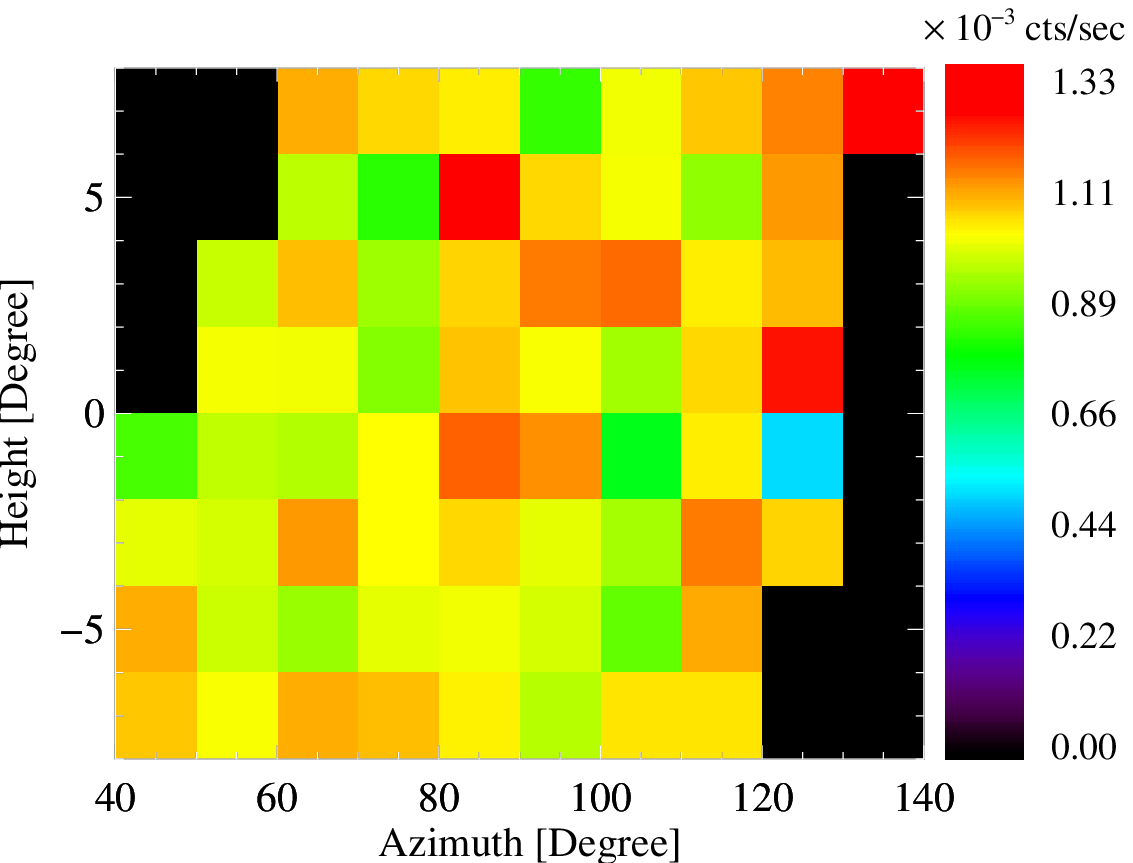}}
    \end{minipage}
    \begin{minipage}{0.49\textwidth}
      \centerline{\includegraphics[width=1.\textwidth]{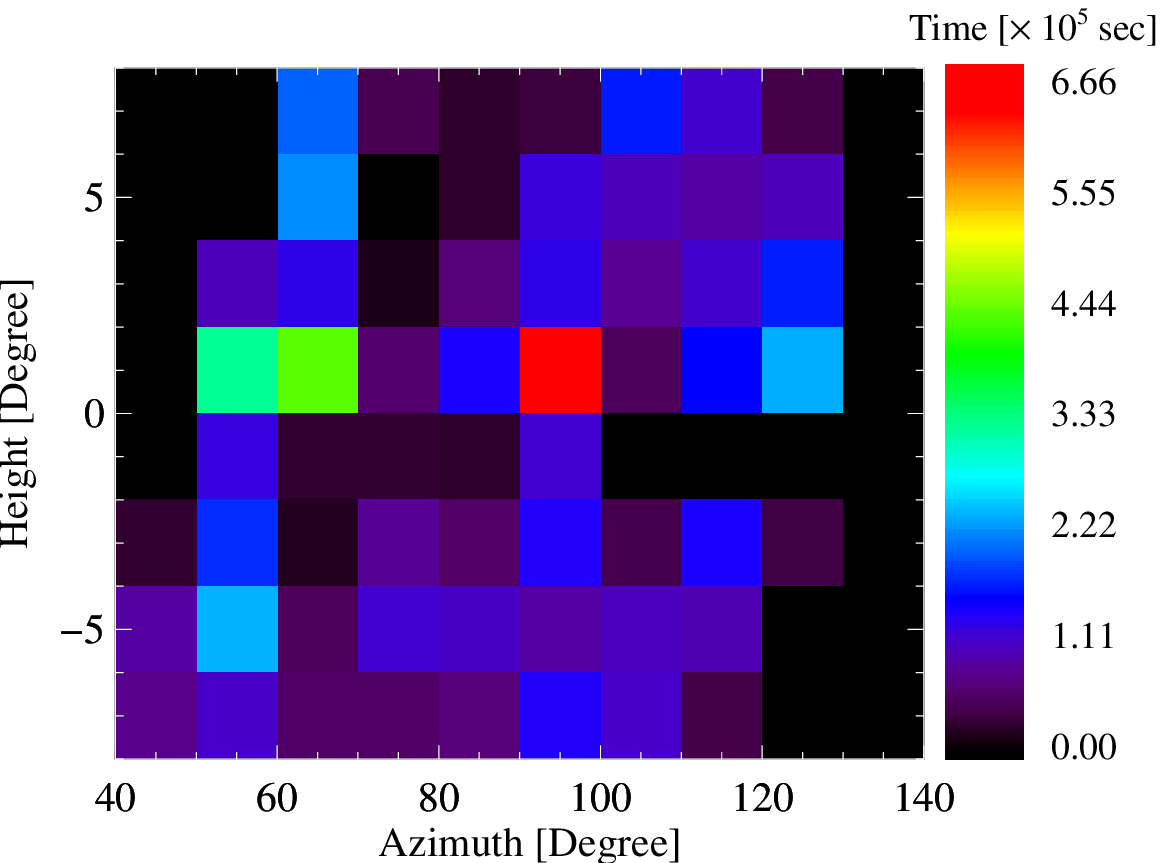}}
    \end{minipage}
  \end{center}
  \caption{\label{fig:background-position} Left: Count rate observed
    with the X-ray telescope depending on the pointing direction of the
    magnet. Each cell has a dimension of $10\,\degree\times2\,\degree$.
    \redtext{The attitude of the X-ray telescope is given in topocentric
      horizontal coordinates.}
    Right: Integration time for each cell.}
\end{figure}

During solar observations the CAST magnet and the detectors change their
orientation relative to the environment of the experimental area (concrete
walls, cryogenic installation). This movement might influence the
background observed by the detectors in a systematic way. Especially the
distance between the CCD detector and the concrete wall close to the CCD
detector changes during individual solar observations. In addition, during
the movement of the magnet, while following the track of the sun, the back
side of the detector will face different areas of the wall which change
during the course of the year. Measurements of the environmental gamma
background in the CAST experimental area with a germanium gamma
spectrometer show a variation of the contribution of, e.g., the
$^{238}\text{U}$ chain to the total environmental background by more than
one order of magnitude between different locations in the experimental
area~\cite{dumont:04a}. To minimize the influence of this effect on the
detector background of the CCD, a lead shield was installed behind the CCD
detector to reduce the apparent background variations below the limit of
sensitivity of the CCD detector.  \Fref{fig:background-position} shows the
background count rate integrated over the full sensitive area of the CCD
depending on the pointing direction of the magnet as observed with the
X-ray telescope in the axion sensitive energy range (left image) and the
corresponding integration time for each cell (right image). The residual
variations apparent from the picture are within the statistical
uncertainties consistent with a constant background level, taking
especially the short integration time in some cells into account.

\section{Conclusions and Outlook}
\label{sec:conclusions-outlook}
The X-ray telescope of CAST is in operation since summer 2003 and is taking
data in routine operation. First results of data acquired in 2003 and 2004
have already been published \citeasnoun{zioutas:05a} and
\citeasnoun{andriamonje:07a}. These results demonstrate how the sensitivity
of rare event experiments as CAST can be improved by a combination of a
focusing optics and a detector with high spatial resolution.
\redtext{Taking into account that an upper limit on $\gagg$ approximately
  depends on the background count rate according to $\gagg\propto b^{1/8}$,
  the X-ray telescope improves the sensitivity of CAST by a factor of
  $\approx 2$, compared to a non-focusing detector system.} Further
background reduction could be achieved by rebuilding the detector
components from materials selected for radio-purity and with a graded-Z
shield close to the pn-CCD chip which acts as an absorber for low energy
photons ($E<7\,\text{keV}$).

\redtext{During 2003 and 2004 CAST explored the axion mass region up to
  $m_{a}\lesssim 0.02\,\text{\ev}$. In the absence of a significant axion
  signal above background an upper limit on the axion to photon coupling of
  $\gagg< 8.8\times10^{-11}\,\text{GeV}^{-1}$ (95\% CL) could be derived.
  This new limit supersedes the previous astrophysical limits for axion
  masses $m_{a}< 0.02\,\text{\ev}$ from the helium-burning lifetime of HB
  stars \cite{raffelt:96a,raffelt:06a}.  The results from this data taking
  period were published in \cite{zioutas:05a} and \cite{andriamonje:07a}.
  Since mid 2006} the CAST experiment probes for axions with a mass $m_{a}>
0.02\,\text{\ev}$ \redtext{(second phase of CAST). To explore this mass
  range} the axion conversion region has to be filled with a buffer gas to
restore coherence between the axion and photon wave function.  By
systematically changing the pressure inside the magnet bore the mass range
from $0.02$ to $0.8\,\text{\ev}$ can be covered. Due to the vast amount of
pressure settings (approximatly 660) necessary to scan the axion mass
region continuously, the integration time per pressure setting (axion mass)
is limited to $1.5\,\text{h}$ per detector. As a consequence we expect
$\approx 0.20\,\text{counts}/\text{run}$ ($1.5\,\text{hours}$ integration
period) as background contribution.  Further optimization of the shielding
or detector materials would not significantly improve the sensitivity of
the X-ray telescope of CAST.

\ack This work has been performed in the CAST collaboration. We thank our
colleagues for their support. We are in debt with W.~Serber, M.~Di~Marco,
and D.~Greenwald for their help with the CCD calibration, telescope
alignment, and for their engagement during the data taking periods of CAST.
We also would like to express our gratitude to the group of the Laboratorio
de Fisica Nuclear y Altas Energias of the University of Zaragoza for
material radiopurity measurements. Furthermore, the authors acknowledge the
helpful discussions within the network on direct dark matter detection of
the ILIAS integrating activity (Contract number: RII3-CT-2003-506222). This
project was also supported by the Bundesministerium f\"ur Bildung und
Forschung (BMBF) under the grant number 05 CC2EEA/9 and 05 CC1RD1/0. and by
the Virtuelles Institut f\"ur Dunkle Materie und \redtext{Neutrinophysik}
-- VIDMAN.

\section*{References}

\bibliography{mnemonic,xmm,cast,conferences,detback,detector}
\bibliographystyle{jphysicsB}

\end{document}